\documentclass[twocolumn,showpacs,floatfix,prb]{revtex4-1}
\usepackage{graphicx}
\usepackage{float}
\usepackage{dcolumn}
\usepackage{bm}
\usepackage{amssymb}
\usepackage{amsmath}
\usepackage{hyperref} 
\hypersetup{bookmarks=true,unicode=true,colorlinks=true, urlcolor=blue,linkcolor=blue,citecolor=blue,filecolor=blue}
\newcommand{\virgolette}[1]{``#1''}
\usepackage[toc,page]{appendix}
\usepackage{physics}

\begin{document}

\title{Plasmons in topological insulator cylindrical nanowires}

\author{P. Iorio, C.A. Perroni, V. Cataudella}
\address{CNR-SPIN and Physics Department "Ettore Pancini", \\
Universita' degli Studi di Napoli ''Federico II'',
\\ Complesso Universitario Monte S. Angelo, Via Cintia, I-80126 Napoli, Italy}

\begin{abstract}

We present a theoretical analysis of Dirac magneto-plasmons in topological insulator nanowires.  We discuss a cylindrical geometry where Berry phase effects induce the opening of a gap at the neutrality point.  By taking into account surface electron wave functions introduced in previous papers and within the random phase approximation, we provide an analytical form of the dynamic structure factor. Dispersions and spectral weights of Dirac plasmons are studied with varying the radius of the cylinder, the surface doping, and the strength of an external magnetic field.  We show that, at zero surface doping, inter-band  damped plasmon-like excitations form at the surface and survive at low electron surface dopings ($\sim 10^{10} cm^{-2} $).  Then, we point out that the plasmon excitations are sensitive to the Berry phase gap closure when an external magnetic field close to half quantum flux is introduced. Indeed,  a  well-defined magneto-plasmon peak is observed at lower energies upon the application of the magnetic field.  Finally, the increase of the surface doping induces a crossover from damped inter-band to sharp intra-band magneto-plasmons which, as expected for large radii and dopings ($\sim 10^{12} cm^{-2}$), approach the proper limit of a two-dimensional surface. 
 \end{abstract}

\maketitle

\section{Introduction}
Topological insulators (TI) are a new class of materials with many novel properties among which  the bulk-boundary correspondence. In a three dimensional TI,  this correspondence leads to the formation of surface states  protected from backscattering by time reversal symmetry  that are observed in the gap of the bulk. The surface states connect the valence band with the conductive band through the formation of Dirac cones \cite{Book1,Book2}. In the case of $Bi_2Se_3$, the electronic states are particularly simple since there is a  single Dirac cone at $\Gamma$ point of the Brillouin zone and a bulk gap around 0.5 eV \cite{D.Hsieh1,D.Hsieh2,D.Hsieh3,Nature}. 

Due to the sizeable spin-orbit interaction characteristic of TI, spatial and spin degrees of freedom are strongly coupled and, for a closed path, the electron spin acquires a $\pi$ phase, known as Berry phase. In particular, this phase provides an anti-periodic boundary condition for the electron wave-function after a $2\pi$ rotation. Therefore, interesting situations take place when TI nano-structures, such as nanowires, are realized.  For a TI nanowire, the Berry phase opens a small gap at the $\Gamma$ point, hampering the formation of the Dirac cone and promoting the formation of different energy sub-bands corresponding to surface states \cite{Peng,Hong,Dufouleur}.  The gap gets closed for a nanowire with an infinite cross-section or with a finite cross-section upon the application of magnetic field which reduces the  effects of the Berry phase \cite{Imura,Paolino}. 

The collective plasmon excitations of Dirac electrons in TI have attracted a huge interest for their potential applications in terahertz detectors\cite{Hongbin} and spintronic devices\cite{Dm}. Only recently, Dirac magneto-plasmons were observed in TI showing high frequency tunability in the mid-infrared and terahertz spectral regions \cite{Lupi,Autore,Autore2,Jun,Stauber}. Due to the spin-momentum locking \cite{Marigliano1, Marigliano2}, these plasmons are always coupled to spin waves acquiring a spin character. For this reason, plasmons in TI are also called \virgolette{spin-plasmons} \cite{Raghu,Yi-Ping,Efmkin}. From the experimental point of view, the observation of Dirac plasmons can be hampered by different effects:  i) they cannot be observed directly through an electromagnetic radiation because the moment conservation is prevented from their dispersion law; ii) moreover, the existence of the impurities present in the material can make the contribution of the surface plasmons negligible compared to the bulk one\cite{Dohum}.  In the former case, a solution can be obtained patterning the surface that add an extra-momentum contribution\cite{Grig}, while, in the latter case, a strategy to avoid the impurity problem is to make the ratio between volume and surface the lowest possible as in the case of nano-structures. This represents one of the reasons why Dirac electrons \cite{Bercioux} and plasmons play an important role in nano-structures, such as nanowires \cite{Paolino,Marigliano3,Perroni1} and nanoplates \cite{Siroki}.

The random phase approximation (RPA) for the dielectric function has been used  for gapped Dirac systems in different  dimensions \cite{Anmol,P k,Guinea}, but, as far as we know, not for the cylindrical nanowire geometry. In this work we extend those studies to the case a cylinder exploiting the analytical electron wave functions that,  in previous works \cite{Imura,Paolino}, have been calculated for a TI cylindrical nanowire with a finite radius in the continuum limit. As part of this work, we analyze the system charge response considering the Coulomb interaction between electrons treating the interaction at the level of RPA. We provide an analytical solution of the inverse dielectric function, and,  then,   of the dynamic factor structure, relevant to study the response of the system to electronic scattering processes. The knowledge of these quantities in the limit of high surface density and infinite radius (two-dimensional surface) is used as a benchmark for the theoretical calculations.

The analysis starts from the case of zero surface doping, i.e. when the chemical potential is inside the small gap opened in the spectrum of the surface states by the effects of the Berry phase. In this case, it is possible to observe the dispersion of damped inter-band plasmon-like excitations that exhibits a minimum threshold value of the momentum along the cylinder axis. Very interesting results are obtained when an applied longitudinal magnetic field closes the gap at half quantum flux and removes the degeneracy of the states by splitting the nanowire sub-bands. As effect of this field, in the dynamic structure factor, we observe a well-defined magneto-plasmon peak with a long life-time and an another excitation peak, resulting from the splitting of the same sub-band, at higher frequency and with lower spectral weight. When the chemical potential crosses the first sub-bands, it is  possible to observe the existence of both damped inter-band and sharp intra-band plasmon excitations with very different frequencies. With increasing the electron doping, the most important electronic excitations become those of intra-band nature. Moreover, the increase of doping or of the cylinder radius induces a cross-over to the two-dimensional regime where the effects of the magnetic field are no more relevant. 

The paper is divided as follows. In Section 2, a low energy continuum model for a $Bi_2Se_3$ cylindrical nanowire is proposed; in Section 3, an analytical form of the dynamic factor structure is provided; in Section 4, inter-band plasmon-like excitations are analyzed for zero surface doping; in Section 5, the case of a finite surface doping is discussed; in Section 6, the features of inter- and intra-band plasmons are compared focusing on their different lifetimes. Details of the calculations are included in Appendix A and B.

\section{The model}
In this section, we recall the single-particle Hamiltonian model, its eigenvalues and eigenstates for a cylindrical wire which represent the starting point for the calculation of the free electron charge susceptibility (or polarization function). Then, within the RPA approximation for the electron-electron Coulomb interaction the dielectric function is obtained. In this section, we focus on the main results since more details of the calculations are provided in Appendices \ref{App:A} and \ref{App:B}. 

\begin{figure}[h!]
\centering
\includegraphics[width=8.5cm]{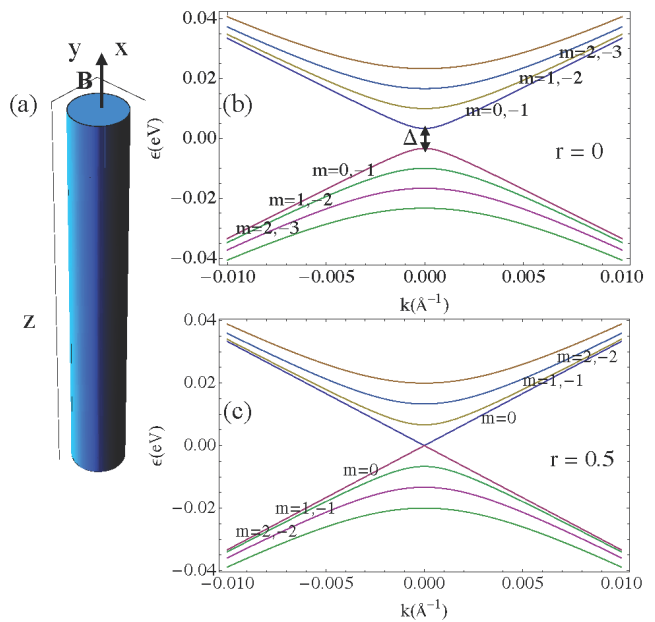}
\caption{\label{fig:1} a) Sketch of the cylindrical wire with translational invariance along $z$ direction in the presence of a magnetic field ${\bf B}$ parallel to the $z$ axis. b) The energy dispersion (in units of eV) as a function of momentum $k$ (in units of $\AA^{-1}$ along z direction with the formation of a gap $\Delta$ and the presence of sub-bands for $r=0$. c) The energy dispersion (in units of eV) as a function of momentum $k$  (in units of $\AA^{-1}$ along z direction with the closure of the gap for $r=0.5$ (half quantum flux). The surface state spectrum is obtained for a cylindrical wire with a radius $R_0=100 \AA$. }
 \end{figure}

As shown in Figure (\ref{fig:1}a), the wire is shaped as a very long cylinder with axis along z direction and radius $R_0$ in the $x-y$ plane.  Following the model proposed in previous works \cite{Paolino,Imura}, the low energy continuum model  describing the electronic properties of $Bi_2Se_3$ close to the $\Gamma$ point is given by the following three dimensional Hamiltonian:

\begin{eqnarray}
\label{eq:0}
H (k) &= & M  \, \mathbb{I}_2 \otimes \, \tau_z + C_1 \,  \sigma_z \otimes \tau_x  k - i C_2 \left[ \sigma_x \otimes \tau_x  (\partial_x+iA_x)  \right.           \nonumber \\
& + & \left. \sigma_y \otimes \tau_x (\partial_y+iA_y) \right].
\label{MainHam}
\end{eqnarray}

In Eq.(\ref{MainHam}), the operator $M = M_0 - M_2 k^2 +M_2 (\partial^2_x+\partial^2_y)$ depends on the parameter $M_0$ controlling the bulk gap, and the parameter $M_2$ giving a mass correction. $\partial_x$ and $\partial_y$ are the partial derivatives along $x$ and $y$, respectively and having assumed the translational invariance along $z$ direction, the quantity $k$ represents the $z$ axis conjugate momentum conserved by the Hamiltonian. The Pauli matrices $\sigma_i$, with $i=x,y,z$, describe the spin degrees of freedom, while the Pauli matrices $\tau_i$, with $i=x,y,z$, take into account the two orbital degrees of freedom of the model. In Eq.(\ref{MainHam}), the symbol $\otimes$ represents the tensor product linking spin and orbital  degrees of freedom. Finally, the parameters $C_2$ and $C_1$ control the inter-orbital and inter-spin couplings, while $A_x$ and $A_y$ are the $x$ and $y$ component, respectively, of the vector potential ${\bf A}$ corresponding to an homogeneous magnetic field ${\bf B}$ parallel to the axis of the cylinder.

In the absence of magnetic field, the Hamiltonian is represented by the following matrix operator:
\begin{equation}
H(k)=
\begin{bmatrix} 
\label{eq:0}
  M          &   C_2 \, k      &       0               &      -i C_1 \partial_- \\ 
  C_2 \, k &   -M              &      -i C_1 \partial_-     &         0       \\
  0            &     -i C_1 \partial_+   &       M              &         -C_2\, k\\
  -i C_1 \partial_+   &     0            &        -C_2\,k      &          -M
\end{bmatrix},
\end{equation}
with the derivative operator $\partial_{\pm}=\partial_x \pm i \partial_y $. The following parameters for $Bi_2Se_3$ are used  in this paper  \cite{Nature,Paolino}: $M_0=-0.28 \,eV$, $M_2=40.00 \, eV \AA^2$  and $C_1=C_2=C=3.33\, eV \AA$. The bulk gap is a few tenths  of $eV$ being proportional to $M_0$. The choice $C_1=C_2$ ensures that the model has particle-hole (p-h) symmetry.

The single-particle energies and states of the Hamiltonian (\ref{MainHam}) can be analytically derived for a cylinder with a large radius (see Figure (\ref{fig:1}a) for a sketch of the system) by using the cylindrical coordinates   $(\rho,\phi,z)$ for the electron position $\vec{r}$  \cite{Paolino,Imura}. We have shown in a previous paper \cite{Paolino}  that, in a cylinder with a radius $R_0$ larger than $100 \AA$,  the surface eigenvalues $\epsilon_{\alpha}$ and eigenfunctions $\psi_{\alpha}(\rho,\phi,z)$ of Eq.(\ref{MainHam}) can be very accurately expressed in the following form:

\begin{equation}
\epsilon_{\alpha} = \epsilon^{s}_{k,m} = s C_2 \sqrt{ k^2+(1+2 m-2 r)^2 \tilde{\Delta}^2(R_0)},
\label{eq:1}
\end{equation}
and
\begin{eqnarray}
\label{eq:2}
\psi_{\alpha}(\rho,\phi,z) &=&\psi_{k,m}^{s}(\rho,\phi,z)\nonumber\\
& =&\sqrt{\frac{1}{4 \pi L}}e^{i k z}R(\rho) e^{i m \phi}\mathbf{u}_{k,m}^{s}(\phi).
\end{eqnarray}

In Eqs.(\ref{eq:1}) and (\ref{eq:2}),  the eigenvalue and eigenstate label is indicated by $\alpha=(k,m,s)$, where the relative number $m$ is related to the symmetry of the cylindrical confinement and $s=\pm$. The effects of a longitudinal magnetic field are included in Eq.(\ref{eq:1}) through the parameter $r$ which is equal to the ratio between the magnetic flux $\Phi$ threaded by the cylinder and the quantum flux $\Phi_0=\hbar/2e$: $r=\Phi/\Phi_0$. 

As reported in Eq.(\ref{eq:1}), a gap $\Delta(R_0)= 2 C_2 \tilde{\Delta}(R_0)$, with $\tilde{\Delta}(R_0)= C_1/(2C_2 R_0)$, opens at $k=0$ in the spectrum of the surface states for $r=0$ (absence of the magnetic field). For a  radius $R=500 \AA$, according to the model parameters considered in in this paper, the surface states have a gap $\Delta=6.66 \, meV$, which is much smaller than the bulk gap. The extra term responsible for the nanowire gap even for $m=0$ is a direct consequence of the Berry phase. As shown in Figure (\ref{fig:1}b), for $r=0$, above and below the semi-gap, the spectrum shows different sub-bands depending on the quantum number $m$. Furthermore for both positive and negative energies, the sub-bands are two-fold degenerate in $m$. For example, in Eq.(\ref{eq:1}) for $r=0$, the couple $m=0$, $m=-1$ corresponds to the same energy. 

We stress that  the case of a wire is quite different from that of a slab. In fact, in the second case, the gap of the system depends only on the thickness of the slab. When the slab is few tens of angstrom thick, the surface states of the two faces interfere with each other  opening of a gap\cite{Xia}. If the thickness is about 
$70 \, \AA$, the Dirac cone is restored closing the gap. Different situation occurs when, regardless of the mutual distance, the two surfaces are always connected to each other, creating a closed loop, as in the case of the wire cross-section. In fact, in this case, we can imagine in a naive way that surface states always communicate with each other (quantum interference effects) through a purely geometric phase, the Berry phase. As a consequence, the opening of a small gap at the $\Gamma$ point occurs even for large nanowire cross-sections.  For the radius considered in this paper, from $500$ to $4000 \, \AA$, the opening of the gap is purely consequence of the Berry phase, and it cannot be ascribed, for example, to the superposition of states on opposite sides of a cylinder diameter. 

In the presence of a longitudinal magnetic field, the gap $\Delta$ gets reduced and the two-fold degeneracy in $m$  removed. Actually, interference terms due to magnetic field weaken the effects of the Berry phase.   In particular, in the case of $r=0.5$ (half quantum flux), a total cancellation of the Berry phase effects takes place because of the magnetic field even for a finite radius nanowire.  For $r=0.5$, as reported in Eq.(\ref{eq:1}), the $m=0$ sub-band shows no more gap. As shown in Figure (\ref{fig:1}c), in the case of $r=0.5$, a two-fold degeneracy is again restored even if it has a different character. For example, in contrast with the case of $r=0$, $m=1$, $m=-1$ are degenerate for $r=0.5$.

In Eq.(\ref{eq:2}),  $L$ is the length of the cylinder along the $z$ axis, $R(\rho)$ is the radial function different from zero for $0<\rho<R_0$, $\mathbf{u}_{k,m}^{s}(\phi)$ is a quadri-spinor  depending on the angular variable $\phi$ (see Appendix \ref{App:A} for details about the function $R(\rho)$ and the four components of $\mathbf{u}_{k,m}^{s}(\phi)$). In the limit of a large radius $R_0$, the radial function $R(\rho)$ for the surface states is essentially localized for values of $\rho$ close to $R_0$.

Using surface single-particle energies and states labelled by $\alpha$, one can calculate the free electron susceptibility $\chi(\vec{r},\vec{r}^\prime;\omega)$ (or polarization function) at zero temperature:
\begin{equation}
\begin{split}
\label{eq:3}
\chi(\vec{r},\vec{r'};\omega) &=  \sum_{\alpha,\alpha^\prime}
\frac{f(\epsilon_{\alpha^\prime})-f(\epsilon_\alpha)}{\epsilon_{\alpha^\prime}-\epsilon_\alpha+\hbar(\omega+i0^+)}\\ 
&\times \; \psi^*_{\alpha^\prime}(\vec{r}) \psi_\alpha(\vec{ r}) \psi^*_\alpha(\vec{ r^\prime}) \psi_{\alpha^\prime}(\vec{r^\prime}),
\end{split}
\end{equation}
where $f(\epsilon_{\alpha})$ is the Fermi function at zero temperature calculated at the energy $\epsilon_{\alpha}$, which is related to the chemical potential $\mu$. 
In the case of the TI nanowire, we use the eigenvalues (\ref{eq:1}) and the corresponding eigenfunctions (\ref{eq:2}) in cylindrical coordinates. Hence, in eq.(\ref{eq:3}),
$\alpha=(k,m,s=\pm)$, $\alpha^\prime=(k+q,m+l,s^\prime=\mp)$, where $q$ is the transferred momentum along $z$ axis, $l$ is the relative number linking the sub-band  $m$ to the  sub-band $m^\prime=m+l$. Indeed, $m$ and $m^\prime=m+l$ are the  variables conjugate to $\phi$ and $\phi^\prime$, respectively. 
The susceptibility for the cylinder becomes
\begin{equation}
\begin{split}
\label{eq:4}
\chi(\vec{r},\vec{r^\prime};\omega) &= \chi(\rho,\rho^\prime,z-z^\prime,\phi-\phi^\prime;\omega)=\\
& R^2(\rho) R^2(\rho^\prime) \chi_0(z-z^\prime,\phi-\phi^\prime;\omega),
\end{split} 
\end{equation}
where $\chi_0(z-z^\prime,\phi-\phi^\prime;\omega)$ reflects the translation invariance along $z$ axis and the rotation invariance around $z$ axis due to the cylindrical geometry (details about $\chi_0(z-z^\prime,\phi-\phi^\prime;\omega)$ are provided in Appendix (\ref{App:A}). Clearly, the susceptibility is different from zero for values of the positions $\vec{r}$ and $\vec{r^\prime}$ inside the cylinder. We point out that, along the radial direction, the susceptibility depends on the product of the squares of the radial functions calculated in $\rho$ and $\rho^\prime$, respectively. This separation of the radial dependences will be of paramount importance for the calculation of the inverse of the dielectric function.
  
Finally, we calculate the dielectric function due to the surface charges of the TI cylindrical wire within the RPA approximation for the electron-electron Coulomb interaction. Even if we confine electrons within the cylinder, the generated electromagnetic fields affects all the space and, therefore, the dielectric function is defined also outside the cylinder. Within the RPA approximation,  the dielectric constant 
$\epsilon(\vec{r},\vec{r^\prime};\omega)$ is defined  in the following way:
 \begin{equation}
 \label{eq:6}
 \epsilon(\vec{r},\vec{r^\prime};\omega)= \delta(\vec{r}-\vec{r^\prime})- \int d\vec{r}_1  V(\vec{r}-\vec{r_1})  \chi(\vec{r_1},\vec{r^\prime};\omega),
 \end{equation}
 where $ V(\vec{r}-\vec{r_1})=e^2/|\vec{r}-\vec{r_1}|$ is the Coulomb potential in real space and the integration is over the volume enclosed by the cylinder \cite{Delerue}. We emphasize again that  $\vec{r}$  can assume values both inside and outside the cylinder unlike $\vec{r_1}$ and  $\vec{r^\prime}$.  In the next section, we will see that, in order to describe the response of the system to external probes, we need to calculate the inverse of the dielectric function $ \epsilon^{-1}(\vec{r},\vec{r^\prime};\omega)$ . As discussed in Appendix (\ref{App:B}), in order to make the inversion of the dielectric function in Eq.(\ref{eq:6}), we solve analytically an integral equation with a separable variable kernel.

\section{Dynamic structure factor}
In this section, we provide an analytic form of the dynamic structure factor for a TI cylindrical wire. This quantity is relevant for example in electron-energy-loss experiments (EELS), where an electron impinges on the sample and loses energy by exciting plasmons. This energy loss is given by the imaginary part of the integral of the potential created by the electron, and the induced charge \cite{Onida}. When the potential due to an electron is taken proportional to a plane wave, one can calculate the response function $L({\vec q}; \omega)$ making a double Fourier transform of the inverse dielectric function:
 
\begin{equation}
\label{eq:7}
L({\vec q}; \omega)=-Im\left[\frac{1}{V}\int \int d\vec{r} d\vec{r^\prime} \epsilon^{-1}(\vec{r},\vec{r^\prime};\omega) e^{-i \vec{q} \cdot \vec{r}} e^{i \vec{q} \cdot \vec{r^\prime}} \right],
\end{equation}
where $V$ is the volume enclosed by the cylinder and  the two integrals are over all the space.  It can be shown that the response function $L({\vec q}; \omega)$ in Eq.(\ref{eq:7}) is proportional to the dynamic structure factor \cite{Friedhelm}, whose peaks characterize the electronic excitations, such as plasmons, induced by scattering processes. Since the system  has translational invariance in the $z$ direction and rotational invariance around $z$ axis, as discussed in Appendix (\ref{App:B}), one can use partial Fourier transforms in $q$ and $l$ for the Coulomb potential $ V(\vec{r}-\vec{r_1})$ and the polarization function $\chi(\vec{r}_1,\vec{r^\prime})$ obtaining the following result:  

\begin{eqnarray}
\label{eq:8}
&& L(q,|{\mathbf q_{||}}|; \omega)= -\frac{2}{R_0^2} Im \biggl [ \sum_l  \frac{\chi_0(q,l;\omega)}{1-\chi_0(q,l;\omega) {\tilde V}_{q,l}} \\
&&\times  \int_{0}^{R_0} d\rho^\prime \rho^\prime \,  {\rm J}_l(|{\mathbf q_{||}}|\rho^\prime) R(\rho^\prime)^2 \int_0^{\infty} d\rho \rho\, {\rm J}_l(|{\mathbf q_{||}}|\rho) S_{q,l}(\rho)\biggr]. \nonumber
\end{eqnarray}
where $|{\mathbf q_{||}}|$ is the modulus  of the momentum transverse to the axis $z$ of the cylinder (parallel to the $x-y$ plane), and ${\rm J}_l$ are the Bessel functions of the first kind. In  Eq.(\ref{eq:8}), $ {\tilde V}_{q,l}$ takes into account the Coulomb repulsion:
\begin{equation}
 {\tilde V}_{q,l}=4 \pi \int d\rho \rho R^2(\rho)  S_{q,l}(\rho),
\end{equation}
since $S_{q,l}(\rho)$ is defined as
\begin{equation}
S_{q,l}(\rho)= e^2 \int d\rho^\prime \rho^\prime  R^2(\rho^\prime)  I_{l}(|q| \rho_<)K_{l}(|q| \rho_>),
\label{eq:8bis}
\end{equation}
with $I_l$ and $K_l$ the modified Bessel functions of the first and second kind, respectively. In  Appendix (\ref{App:A}) we report the expansion of the Coulomb potential in cylindrical coordinates clarifying that $ \rho_<$ and $\rho_>$ appearing in Eq.(\ref{eq:8bis}) represent the smaller and larger of $\rho$ and
$\rho^\prime$, respectively.

The zeros of the denominator in equation (\ref{eq:8}) provide the plasmon dispersions. Indeed, one has  the equation
\begin{equation}
1- \chi_0(q,l;\omega)  {\tilde V}_{q,l}=0
\label{zero}
\end{equation}
 typical of the RPA approach.  In this paper, we focus on the contributions for $|{\mathbf q_{||}}| \rightarrow 0$, therefore we consider probes  propagating parallel to the axis of the cylinder. This means that we only have to analyze the term $L(q, 0;\omega)$ corresponding to $l=0$:
\begin{equation}
\label{eq:9}
 L(q, 0;\omega)= -\frac{2}{R_0^2} Im \biggl [ \frac{\chi_0(q,0;\omega)}{1-\chi_0(q,0;\omega)  {\tilde V}_{q,0}}\int_0^{\infty} d\rho \rho\, S_{q,0}(\rho)\biggr]. 
\end{equation}

In the limit of large radius $R_0$ and small $q$, it is possible to show how the potential $R_0 {\tilde V_{q,0}}$ converges to the two-dimensional  Coulomb potential 
$V_q^{2D}= 2 \pi e^2/q$.  As reported in Appendix \ref{App:A}, in the limit of infinite radius,  the dynamic polarization $\chi_0(q,0;\omega)$ also converges to that of the two-dimensional case \cite{Guinea} . 

In addition to the plasmon frequencies and their spectral weights, we can also calculate the induced charge density $d (\vec{r})$, potential $\Phi(\vec{r})$ and electric field  
$\vec{E}(\vec{r})$ corresponding to these modes. In Appendix \ref{App:A}, we provide the analytical expressions of these quantitities and all the details of the calculations. We show that the electronic charge density corresponding to a mode at fixed $q$ and $l$ is
\begin{equation}
\label{eq:rhoo}
d (\vec{r}) =d(\rho,\phi,z) =e^{i q z} e^{i l \phi} B_{q,l}(R_0)  R^2(\rho),
\end{equation}
where the dependence on $\rho$ is only through the square of the radial function $R(\rho)$, and $B_{q,l}(R_0)$ is an arbitrary constant depending in general also on the cylinder radius $R_0$. As shown in Appendix \ref{App:A}, through the density, one can calculate the electric potential $\Phi$ relative to a mode at fixed $q$ and $l$:
\begin{eqnarray}
\label{eq:poten}
\Phi(\vec{r})=\Phi(\rho,\phi,z)&= & e^{i q z}e^{i l \phi} B_{q,l}(R_0) \times \\ 
&& \int_{0}^{R_0} d\rho^{\prime}\rho^{\prime} R^2(\rho^{\prime}) K_l(q \rho_>) I_l(q \rho_<). \nonumber
\end{eqnarray}

Through the potential, one can determine the electric field $\vec{E}(\vec{r})=-\nabla \Phi(\vec{r})$ inside ($\rho<R_0$) and outside ($\rho>R_0$) the cylinder. In principle, all the three components exist: radial $E_{\rho}$, longitudinal $E_z$, and angular $E_{\phi}$. We note immediately that if we take only the modes at $l=0$, the dependence on the angular part disappears in Eq. (\ref{eq:poten}). Therefore, for the modes at $l=0$, there are only longitudinal and transverse components  of the field, while, for modes with $l$ different from zero, also the angular component $E_{\phi}$ appears. As discussed in Appendix \ref{App:A}, the electric field for $l=0$ is similar to that of the $TM$ mode at $l=0$  in the case of surface plasmons for conventional metallic cylinders \cite{sarid}.  In this paper,  we analyze the electrostatic limit, where the magnetic field is always assumed zero. Therefore, the axial component of the magnetic field is clearly zero for the $TM$ mode at $l=0$. We recall that, in TI, Hall currents can lead to a magneto-electric effect according to which a magnetic field can induce a charge polarization. As a consequence, the separation between TE and TM is not exact in TI cylinders for l=0  \cite{schutky,LiCui}. However, the magneto-electric effect can be treated at a perturbative level and have practically no impact on the plasmon properties of TI systems.
We remark that, within the electrostatic limit adopted in this paper, these very weak effects on the plasmon are automatically neglected. 

As discussed in Appendix \ref{App:A}, for the radii chosen in this paper ($R_0 \ge 500 \AA$), the induced density becomes practically different from zero only on the lateral surface of the cylinder. For the modes with $l=0$, we can provide analytical expressions for the electric field: $(E_{\rho}, E_z)= B_{q,l=0}(R_0) \times $  
\begin{equation}
\label{field}
 \left\{
        \begin{array}{ll}
          \left(q I_1(q \rho) K_0(q R_0) e^{i q z}, \, i q  I_0(q \rho) K_0(q R_0) e^{i q z} \right) , \ \  \rho \le R_0, \\
           \left( -q K_1(q \rho) I_0(q R_0) e^{i q z}, i q K_0(q \rho) I_0(q R_0) e^{i q z} \right) , \ \  \rho>R_0.
        \end{array}
    \right.
\end{equation}
Inside the cylinder ($\rho<R_0$),  the  radial component $E_{\rho}$ goes as $ I_1(q r)$, therefore, at finite $q$, $E_{\rho}$ is proportional to $q \rho$ for very small values of $\rho$.  On the other hand, outside the cylinder ($\rho>R_0$), $E_\rho$ goes as $K_1(q \rho)$, therefore, at finite $q$, $E_\rho$ is proportional to 
$ e^{-q \rho}/ \sqrt{q \rho} $ for very large values of $\rho$. Moreover, as expected, for $q \rightarrow 0$, the electric field inside the cylinder is zero, while, outside the cylinder, the longitudinal component $E_z$ vanishes, and the radial component $E_{\rho}$ goes as $1/\rho$.
It is apparent that the long-range spatial variation of the electric field is completely different from that of the charge density which is essentially localized on the lateral surface of the cylinder.

For a system with p-h symmetry, $\chi_0(q,l;\omega)=\chi_0(q,l;-\omega)^*$. Therefore, the results will be given only for chemical potential $\mu>0$, $\omega>0$ and $q>0$. Furthermore, since we will consider only the $l=0$ contribution, in the following, we name $ {\tilde V}_{q,0}= {\tilde V}_q$ and $\chi_0(q,0;\omega)=\chi_0(q;\omega)$. As shown in  Appendix \ref{App:A} and following what done in literature, we can divide the polarization function for massive Dirac electrons into three contributions:
\begin{equation}
\chi(q;\omega)=-\chi_{\infty}^{-}(q;\omega)+\chi_{\mu}^{-}(q;\omega)+\chi_{\mu}^{+}(q;\omega),
\label{eq:39}
\end{equation}
where the subscripts are related to the position of the chemical potential: $-\chi_{\infty}^{-}(q;\omega)$ is the contribution for $\mu<\Delta$, $\chi_{\mu}^{-}(q;\omega)+\chi_{\mu}^{+}(q;\omega)$ is the additional contribution for $\mu>\Delta$. In Eq.(\ref{eq:39}), the superscript $+$ indicates intra-band transitions, while the superscript $-$ interband transitions. Therefore, one can define
\begin{equation}
\begin{split}
\label{eq:40}
\chi^\pm_{T}(q;\omega)&=\frac{1}{4 C_2 \pi^2} \sum_m \int_{-T}^{T} dk \biggl[1\pm   \frac{k (k+q)+\tilde{\Delta}^2_m(R_0)}
{\tilde{\epsilon}_{k,m}\tilde{\epsilon}_{k+q,m}} \biggr] \\
&\times \left[ \frac{\tilde{\epsilon}_{k,m} \mp \tilde{\epsilon}_{k+q,m}}{(\tilde{\omega}+i \tilde{\eta})^2-(\tilde{\epsilon}_{k,m}\mp \tilde{\epsilon}_{k+q,m})^2} \right],
\end{split}
\end{equation}
with $\tilde{\Delta}_m(R_0)=\tilde{\Delta}(R_0)(1+2m-2 r)$, $\tilde{\omega}=\hbar \omega/C_2$, $\tilde{\epsilon}_{k,m}=\epsilon_{k,m}/C_2$, $\tilde{\mu}=\mu/C_2$, and $\tilde{\eta}=\eta/C_2$, with $\eta$ very small positive energy. The integration limit $T$ is equal to infinity for $\mu$ in gap, while $T=\sqrt{\tilde{\mu}^2-(1+2m-2r)^2 \tilde{\Delta}^2}$ otherwise. The sum over $m$ for the case $\mu<\Delta$ has to be made over all sub-bands, then $m$ goes from minus infinity to infinity. On the other hand, for $\mu>\Delta$, the sum is limited by the condition of existence of $T$, that is 
$\tilde{\mu}^2-\tilde{\Delta}^2(1+2m-2 r)^2>0$, with $m$ integer.

\section{Zero surface doping}

In a previous section, we have pointed out that the effect of the Berry phase is to open a small gap in the energy spectrum. In this section, we provide a description of the inter-band surface excitations, when the chemical potential $\mu$ is in the Berry phase gap. This means that the chemical potential is in the middle of the gap: $\mu=0$ (due to the p-h symmetry). As shown  in Figure (\ref{fig:1}), the basic p-h excitations are due to transitions from states with negative energies ($s=-$ in Eq.(\ref{eq:1})) to those with positive energies ($s=+$ in Eq.(\ref{eq:1})). 

As discussed in this section, in the absence of an applied magnetic field, the inter-band excitations are severely damped. Moreover, these inter-band excitations are not present in the limit $q \rightarrow 0$. For this reason, we call the inter-band mode a plasmon-like excitation. This will help comparing this excitation with the sharp intra-band mode which is well-defined in   the limit $q \rightarrow 0$. In the following, the intra-band mode will be called intra-band plasmon. In the next two sections, we will analyze the features of inter-band and intra-band excitations, respectively.  In Section 6, the short lifetimes of inter-band plasmon-like excitations  will be compared with long ones of intra-band plasmons.

\begin{figure}[h!]
\centering
\includegraphics[width=8.5cm]{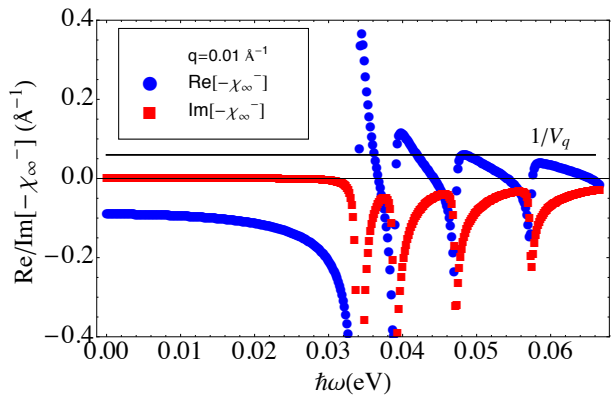}
\caption{Real (blue circles) and imaginary (red squares) part of the susceptibility (in units of  $\AA^{-1}$, obtained numerically with $\eta=1.5 \times10^{-4} \; eV$) as a function of the energy (in units of eV) for $q=0.01 \; \AA^{-1}$  at $R_0=500 \AA$ and $r=0$. Large variations of the real part are related to those of the imaginary part in correspondence with the same value of $\hbar \omega=C_2\sqrt{q^2+4 \tilde{\Delta}^2 (1+2m-2 r)^2}$ at $r=0$. The intersections that $1/ {\tilde V}_q$ (black line) has with $\Re( - \chi_{\infty}^-)$ give the plasmon frequencies.}
 \label{fig:2}
 \end{figure}

In this section, considering the chemical potential in gap ($\mu=0 < \Delta$), we need only to analyze the contribution $-\chi_{\infty}^-$ of the equation (\ref{eq:39}). Moreover, the $l=0$ contribution to the susceptibility come only from inter-band transitions associated to the same sub-band number $m$ but opposite $s$ in Eq.(\ref{eq:1}) (see for clarity also Fig.(\ref{fig:1})).  

\begin{figure}[h!]
\centering
\includegraphics[width=8.5cm]{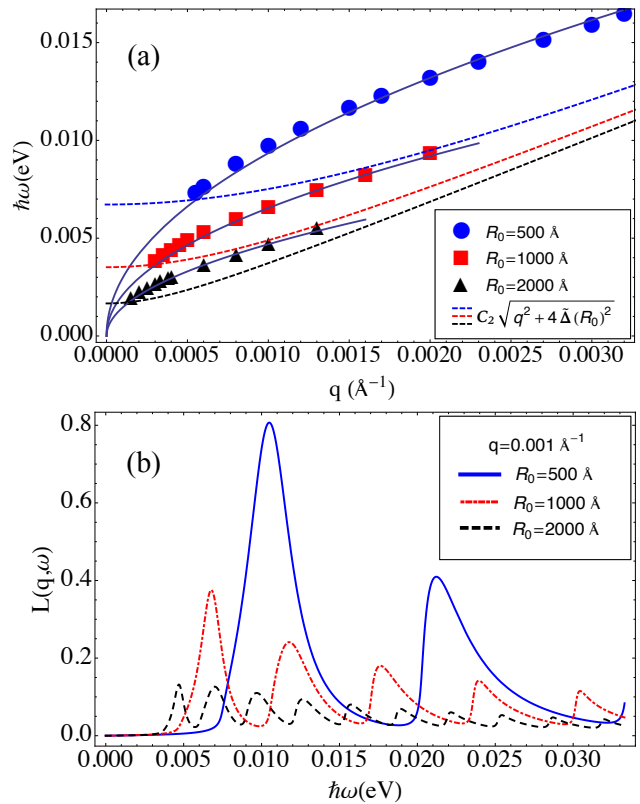}
\caption{a) Inter-band plasmon-like dispersion  (in units of eV) as a function of the wave-vector $q$ (in units of $\AA^{-1}$) for different values of the cylinder radius. Plasmon excitations are present only for energies larger than $C_2 \sqrt{q^2+4\tilde{\Delta}^2(R_0)}$. 
The fits (continuous lines) of the dispersions indicate that, above the minimum $q$, the dispersion is  square root like: $\hbar \omega =\alpha  \sqrt q$. For $R_0=500 \AA$, $\alpha \sim 0.3 eV \AA^{1/2} $; for $R_0=1000 \AA$, $\alpha \sim 0.2 eV \AA^{1/2}$ ; for $R_0=2000 \AA$, $\alpha \sim 0.155 eV \AA^{1/2}$ . b) the dynamic structure factor $L(q;\omega)$ as a function of the energy (in units of eV) at $q=0.001 \;\AA^{-1}$ for different values of the cylinder radius.}
\label{fig:3} 
 \end{figure}

We determine the plasmon dispersion from Eq.(\ref{zero}) considering $l=0$: 
\begin{equation}
\label{eq:41}
1- \operatorname{\mathbb{R}e}\left[\chi_0(q;\omega) \right]{\tilde V}_q=0.
\end{equation}
We solve numerically equation (\ref{eq:41}) taking $\eta$ (see Eq. (\ref{eq:40})) very small but finite. As shown in an example in  Figure (\ref{fig:2}), the equation has more solutions corresponding to different frequencies, all with a finite imaginary part. Furthermore, for each value of $R_0$, these inter-band plasmon-like excitations show a minimum frequency related to the finite gap $\Delta(R_0)$. In fact, the real part $\Re(-\chi_{\infty}^-)$ is completely negative for $\hbar \omega<C_2 \sqrt{q^2+4\tilde{\Delta}^2(R_0)}$, instead it assumes also positive values for larger values of frequency. Since the solutions of the equation (\ref{eq:41}) exist only for values of $\Re(-\chi_{\infty}^-)>0$, plasmon dispersion exists only above the curve $\hbar \omega=C_2 \sqrt{q^2+4\tilde{\Delta}^2(R_0)}$. In Fig. (\ref{fig:3}a) we report these curves for different cylinder radii.

Associated with minimum frequency $\omega$, there is also a minimum momentum $q_t$ for which equation (\ref{eq:41}) admits a solution. This is clearly shown in  Figure  (\ref{fig:3}a). With decreasing the radius $R_0$, the gap $\Delta(R_0)$ gets enhanced, therefore this minimum $q_t$ becomes bigger. It is worthwhile noting that, due to the existence of a minimum threshold $q_t$, it is not possible to derive an analytical expression of the dispersion in the limit of small $q$, as in the case of intra-band plasmons.
In Fig. (\ref{fig:3}a) we report our numerical analysis for the plasmon dispersions at different cylinder radii.

It is important to analyze the behavior of the potential $ {\tilde V}_q$ in the limit of small $q$, since two different regimes are present: 
\begin{equation}
\label{eq:42}
 {\tilde V}_q \sim \left\{
        \begin{array}{ll}
            e^2 K_0(q R_0) \mapsto -e^2 \log{(q R_0)}, \ \ & \quad q \ll 1/R_0, \\
             e^2/(q R_0)= V_q^{2D} / R_0 , \ \ & \quad q> 1/R_0.
        \end{array}
    \right.
\end{equation}
Indeed, for $q \ll 1/R_0$, $ {\tilde V}_q$ shows a one-dimensional type behavior, while, for $q> 1/R_0$  it is more two-dimensional \cite{Anmol}. 

Inter-band plasmon-like excitations are stable only for $q>q_t$, which is comparable with $1/R_0$ in the absence of the magnetic field. Therefore, as shown in Figure (\ref{fig:3}a), the plasmon dispersion is similar to the two-dimensional case ($\omega \sim \sqrt{q}$). Moreover, we observe that, for larger values of $q$, the dispersion tends towards $C_2 q$. For the radii considered in this paper, the linear regime is reached at most at $q>0.01\AA^{-1}$. This value of $q$ is in any case smaller than $0.1$ $\AA^{-1}$, which represents the order of magnitude of the wave-vector at the border of the Brillouin zone in materials like $Bi_2Se_3$. Therefore, the analysis pursued in this paper considers values of the wave-vector $q$ where RPA is known to give a reliable description of the electronic excitations.    
The dynamic structure factor calculated in equation (\ref{eq:9}) provides the spectral weights associated with the plasmon frequencies. As shown in Figure (\ref{fig:3}b),
this allows to estimate the damping of the plasmon and its lifetime. In the case of inter-band plasmon-like excitation studied in this section, we stress that, for long wavelength, only the first peak corresponds to a solution of equation (\ref{eq:41}). The peaks at higher frequencies are associated to minima of the right side member of eq.(\ref{eq:41}), and correspond to higher values of the sub-band number $m$. Actually, we notice that only the first peak has a Lorentzian shape while the others become progressively more asymmetric with increasing the frequency. As shown in Figure (\ref{fig:3}b), for the radius $R_0=500 \AA$, the first plasmon peak almost saturates the spectral weight, while, for the radius $R_0=2000 \AA$, it  has a spectral weight comparable with high frequency peaks. In any case, the increase of the radius $R_0$  induces a decrease of the spectral weight  as we see in Figure (\ref{fig:3}b) and a shift of the plasmon frequency toward lower values. If one further increases the value of $R_0$, the spectral weight of the inter-band plasmon frequency falls to zero. Indeed, as discussed in Appendix \ref{App:A}, for $R_0 \rightarrow \infty$, one reaches, as expected, the two-dimensional limit where inter-band plasmon-like excitations are no more present. A similar decrease of the plasmon spectral weight is observed when wave-vector $q$ increases.

\begin{figure}[h!]
\centering
\includegraphics[width=8.5cm]{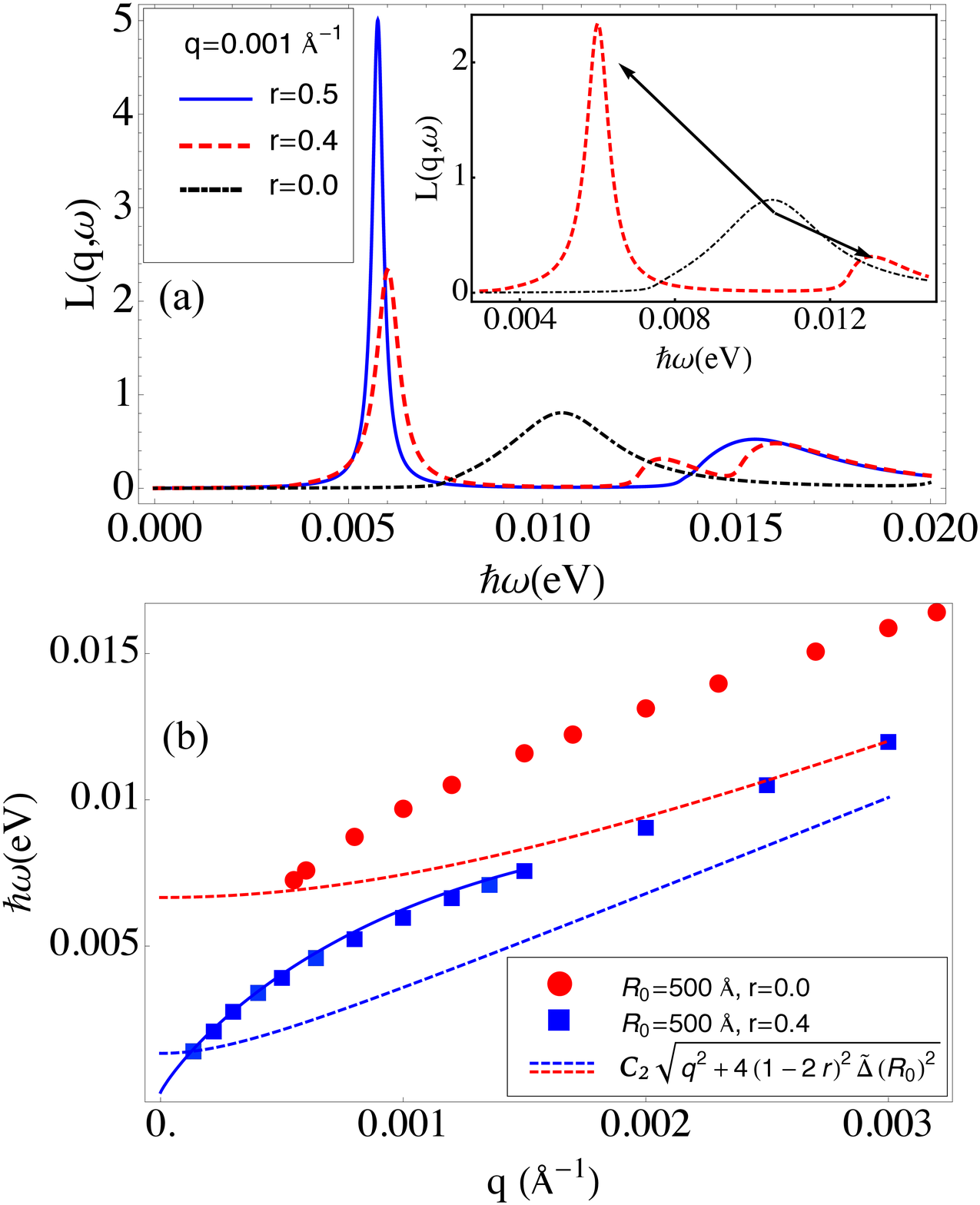}
\caption{ a) The dynamic structure factor as a function of the energy (in units of eV) at $R_0=500 \; \AA$ and $q=0.001 \; \AA^{-1}$ for some values of the ratio of the magnetic fluxes. Note the separation of the plasmon peak into two peaks (shown in the inset by the arrows).  b) Plasmon dispersion (in units of eV) as a function of the momentum $q$ (in units of $\AA^{-1})$ at $R_0=500 \AA$ for two values of $r$: $r=0$ (red circles), $r=0.4$ (blue squares).  Plasmon excitations are present only for energies larger than 
$C_2 \sqrt{q^2+4(1-2r )^2 \tilde{\Delta}^2(R_0)}$. 
The fit (continuous line) of the $r=0.4$ curve provides a dispersion typical of a one-dimensional system: $\hbar \omega=  \alpha q \sqrt{K_0(q R_0)}$, with $\alpha=6.5 eV \AA^{-1}$. }
\label{fig:4}
 \end{figure}

The application of a magnetic field along the axis of the cylindrical wire introduces a magnetic flux $\Phi$ which changes the gap value through the term $r=\Phi/\Phi_0$ present in equation (\ref{eq:1}). This way, it is possible to study the effects of the Berry phase on Dirac inter-band plasmon frequencies in nanowires up to the closing of the gap. As shown in Figure (\ref{fig:1}), the additional effect of the magnetic field is to remove the degeneracy of energy states  and it has an interesting effect on the dynamic structure factor. In fact, the presence of the magnetic field separates a peak into two: one at a frequency lower and the other at a frequency larger than the peak at $r=0$, as shown in Figure (\ref{fig:4}a). 

We notice that the sum of the spectral weights of two new plasmon peaks is similar to that of the main peak for zero magnetic field. The redistribution of this weight is not homogeneous, indeed almost all the contribution is given by the low frequency peak, as we show in the inset of Figure (\ref{fig:4}a). Not only the integral of this peak is more important but also the peak itself is better defined being narrower and higher. Indeed, the closure of the Berry gap caused by the magnetic field gives rise to a magneto-plasmon with a very long life-time, therefore more likely to be observed experimentally. Our analysis suggests that, in the presence of a magnetic field, the magneto-plasmon peak should be observed at lower frequency. Furthermore, for a magnetic field near half quantum flux, as shown in Figure (\ref{fig:4}b) for $r=0.4$, when the gap is practically closed, the minimum threshold $q_t$ becomes very close to zero. As discussed in equation (\ref{eq:42}), the potential for values of $q<<1/R_0$ has a one-dimensional  behaviour, and, therefore, also the magneto-plasmon for wavelength $q_t<q<<1/R_0$  acquires a dispersion typical of one-dimensional systems: 
$\omega \propto q \sqrt{K_0(q R_0)}$, as shown in Figure (\ref{fig:4}b). Unlike the case without magnetic field, a threshold minimum $q_t$ tending to zero with the closing of the gap allows us to explore the region which has a one-dimensional character.
Finally, with the closing of the gap {\em at $r=0.5$}, we restore the degeneration in the sub-bands but the resulting excitations are at frequencies lower than those for zero magnetic field. Indeed, the resulting peak of the dynamic structure factor has a similar spectral weight but it is much higher and narrower. We will see in the next section that the nature of this magneto-plasmon at zero gap is different  from that of inter-band type since it is similar to an intra-band plasmon. 

\begin{figure}[h!]
\centering
\includegraphics[width=9.cm]{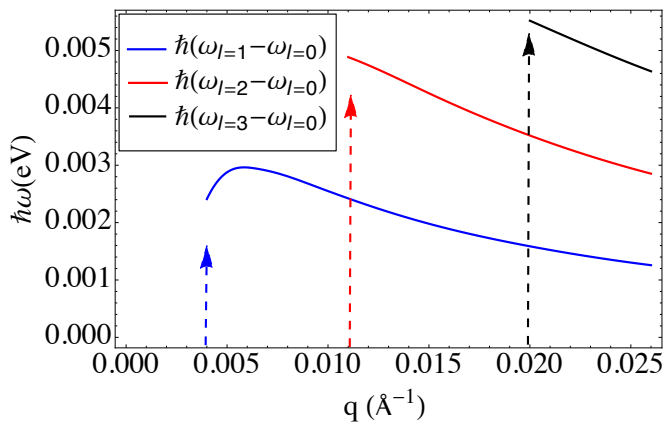}
\caption{ Differences of the frequencies (in units of eV)  at $l=\pm1$, $\pm 2$ and $\pm 3$, respectively, with that at $l=0$  as a function of the wave-vector $q$ (in units of $\AA^{-1}$) for a radius $R_0=500 \,  \AA$ in the case of zero surface doping. The dashed arrows indicate the different minimum thresholds $q_t$ for $l=\pm1$, $\pm 2$ and $\pm 3$, respectively.}  
\label{fig_l_different}
 \end{figure}

Up to now, the entire analysis for inter-band plasmon-like excitation has been made for the mode at $l=0$.  However, the equations (\ref{eq:8}) and (\ref{zero}) depend on a generic relative number $l$. Through the dynamic structure factor, a spectral weight can be associated with each zero of the RPA equation in (\ref{zero}) for different $l$.  A finite value of $l$ means selecting different transitions between the sub-bands shown in Figure (\ref{fig:1}). In fact, for example, $l=\pm 1$ means going from sub-band $m$ to sub-band $m\pm1$. We derive the plasmon dispersion at zero doping for $l=\pm1$, $\pm 2$ and finally $\pm 3$ and, as shown in Fig. (\ref{fig_l_different}), we study the difference of the plasmon frequencies at finite $l$ with respect to the frequencies at $l=0$. Of course, we find the same zero for opposite values of $l$. That's why in Fig. (\ref{fig_l_different}) we report only positive values of $l$. The three dispersions are very close in frequency and have the same monotony, but, with the increase of the parameter $l$, also  the minimum threshold $q_t$ (indicated by the dashed arrows in Fig. (\ref{fig_l_different}))  gets significantly enhanced. This is an expected result since a greater $l$  involves higher energy transitions. Therefore, the required $q_t$ must be higher and comparable to the minimum difference energy between sub-bands. 

For the solutions found in Figure (\ref{fig_l_different}), we can evaluate the spectral weight through the dynamic structure factor of Eq. (\ref{eq:8}).  Only for the conjugate variable $q_{||} \ne 0$, values of $l$ different from zero can be considered.  If one chooses small values of $q_{||}$, the momentum  appears practically parallel to the axis of the cylinder. In particular, for each value of $q$, we have considered a $q_{||}$ about an order of magnitude lower. The result is that, with increasing $l$, the spectral weight decreases significantly, confirming that the contribution for the mode at $l=0$ is the most important. In fact, already for $l=3$, the spectral weight can be considered practically equal zero being three orders of magnitude smaller than that at $l=0$.


\section{Finite surface doping}

In this section we analyze the case of finite surface electron doping: $\mu > \Delta$. In order to find the dispersion and spectral weight associated with the plasmon excitations, we have to consider all three contributions of equation (\ref{eq:39}). We start analyzing the case of low electron doping to understand how the crossover between inter- and intra-band plasmon takes place. 

We define the electron doping for the case of a cylindrical wire as follows:

\begin{equation}
\label{eq:43}
n=\frac{1}{2 \pi^2 R_0} \sum_{m=-M+1}^{M-1} \sqrt{\tilde{\mu}^2-(1+2 m-2 r)^2 \tilde{\Delta}^2},
\end{equation}
where $M$ is the number of the sub-bands intersected by chemical potential.
We choose to study the case in which the potential intersects a single sub-band equivalent to $m = 0$. For this condition we have an electron doping equal to 
$n=10^{10} \; cm^{-2}$.  This way, the equation (\ref{eq:41}) always admits solution for each value of  momentum $q$ due to the intra-band contribution given by $\chi_{\mu}^{+}(q;\omega)$, unlike the case of $\mu=0$ studied in the previous section where there was a minimum threshold $q_t$. For this reason, it is possible to make the $q \rightarrow 0$ limit for the polarization function valid at each $\omega$:
\begin{equation}
\label{eq:44}
\chi_0(q \rightarrow 0;\omega) = \frac{1}{2 \pi^2 C_2} \sum_m \frac{k_F(m) \;\tilde{\mu}\; q^2}{\tilde{\mu} \; \omega^2 - k_F(m) \; q^2  },
\end{equation}
where $k_F(m)=\sqrt{\tilde{\mu}^2-(1+2m-2r)^2 \tilde{\Delta}^2}$ and the sum on $m$  depends always on the number of sub-bands intersected by the chemical potential.  At this point we have all the ingredients to define the intra-band plasmon analyzing the dispersion and the response function. 

\begin{figure}[h!]
\centering
\includegraphics[width=8.5cm]{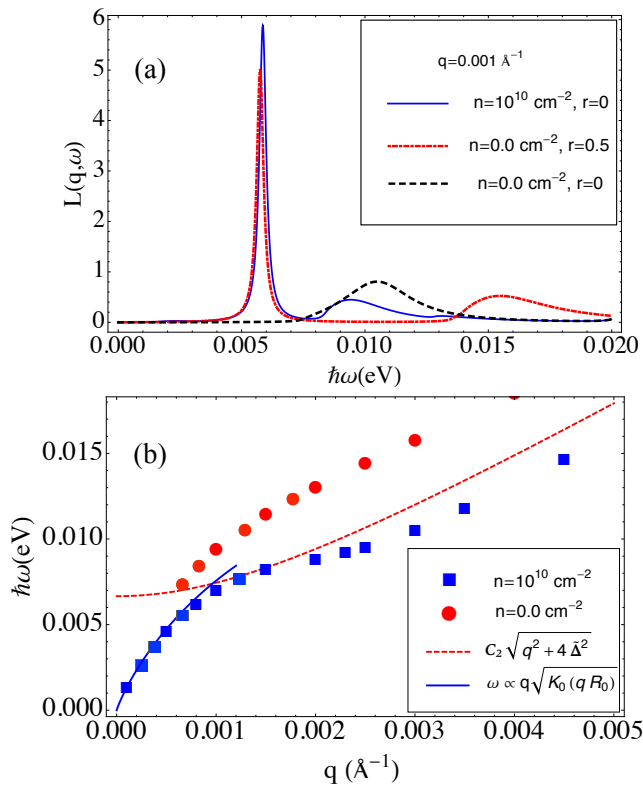}
\caption{ a) The dynamic structure factor as a function of the energy (in units of eV) at $R_0=500 \; \AA$ and $q=0.001 \; \AA^{-1}$ for electronic doping $n=10^{10} \;cm^{-2}$ and $r=0$ (solid blue line), $n=0 \; cm^{-2}$ and $r=0$ (dashed black line), and $r=0.5$ (red dotted line). b) Plasmon dispersion (in units of eV) as a function of the momentum $q$ (in units of $\AA^{-1}$)  in the case of single occupied subband with $n=10^{10} \;cm^{-2}$ (blue squares) compared with the inter-band contribution for zero electronic doping (red circles). Inter-band plasmon-like excitations are present only for energies larger than $C_2 \sqrt{q^2+4(1-2r )^2 \tilde{\Delta}^2(R_0)}$. Note how the analytical solution (continuous line) for small $q$ matches the points obtained numerically.  }
 \label{fig:5} 
 \end{figure}

We start from the case of electron doping $n=10^{10} \; cm^{-2}$ and $R_0=500 \AA$ corresponding to a single occupied sub-band. 
We note immediately that $L(q;\omega)$ (see Figure (\ref{fig:5}a)) exhibits two peaks. The first one at lower energy is well defined with shape, spectral weight and position very similar to those we have  found  for the magneto-plasmon at $\mu=0$ and $r=0.5$. On the contrary, the second peak at higher energy is much broader and similar to the peak associated at the inter-band plasmon-like excitation observed at $\mu=0$ and $r=0$. Actually, the effect of the gap closure in the presence of the magnetic field at $\mu=0$ gives a result very close to what observed for low doping in the absence of the field. Therefore, for low doping (only first subband occupied) we observe that the inter-band Dirac plasmon-like excitation still survives even if its spectral weight is reduced, but a new well defined excitation (intra-band plasmon) gets in.

Finally, to derive the dispersion for long wavelength, we can consider  equation (\ref{eq:44}) and replace it in Eq.(\ref{eq:41}) thus obtaining the behaviors  for the case of single occupied sub-band:
\begin{equation}
\label{eq:44-1}
 \frac{1}{2 \pi^2 C_2}  \frac{k_F(0) \;\tilde{\mu}\; q^2}{\tilde{\mu} \; \omega^2 - k_F(0) \; q^2  } {\tilde V}_q=1,
\end{equation}
which has solution $\omega_p=q\sqrt{C_2^2 k_F^2 \pi +4 k_F C_2 \tilde{\mu}  {\tilde V}_q}/\sqrt{\pi \tilde{\mu}^2}$. We recall that $ {\tilde V}_q$ for $q \rightarrow 0$ has different trends depending on whether $q<<1/R_0$ (one-dimensional) or $q>1/R_0$ (two-dimensional). For a single sub-band, we call $k_F(m=0)=k_F$. Therefore, we find that the  $q \rightarrow 0$ behavior for the low doping dispersion is of one-dimensional type with $\omega_p \propto q\sqrt{K_0(q R_0)}$. In fig.(\ref{fig:5}b) the analytic solution of equation (\ref{eq:44-1}) is shown as the continuous line and shows a very good agreement with the numerical data.

\begin{figure}[h!]
\centering
\includegraphics[width=8.5cm]{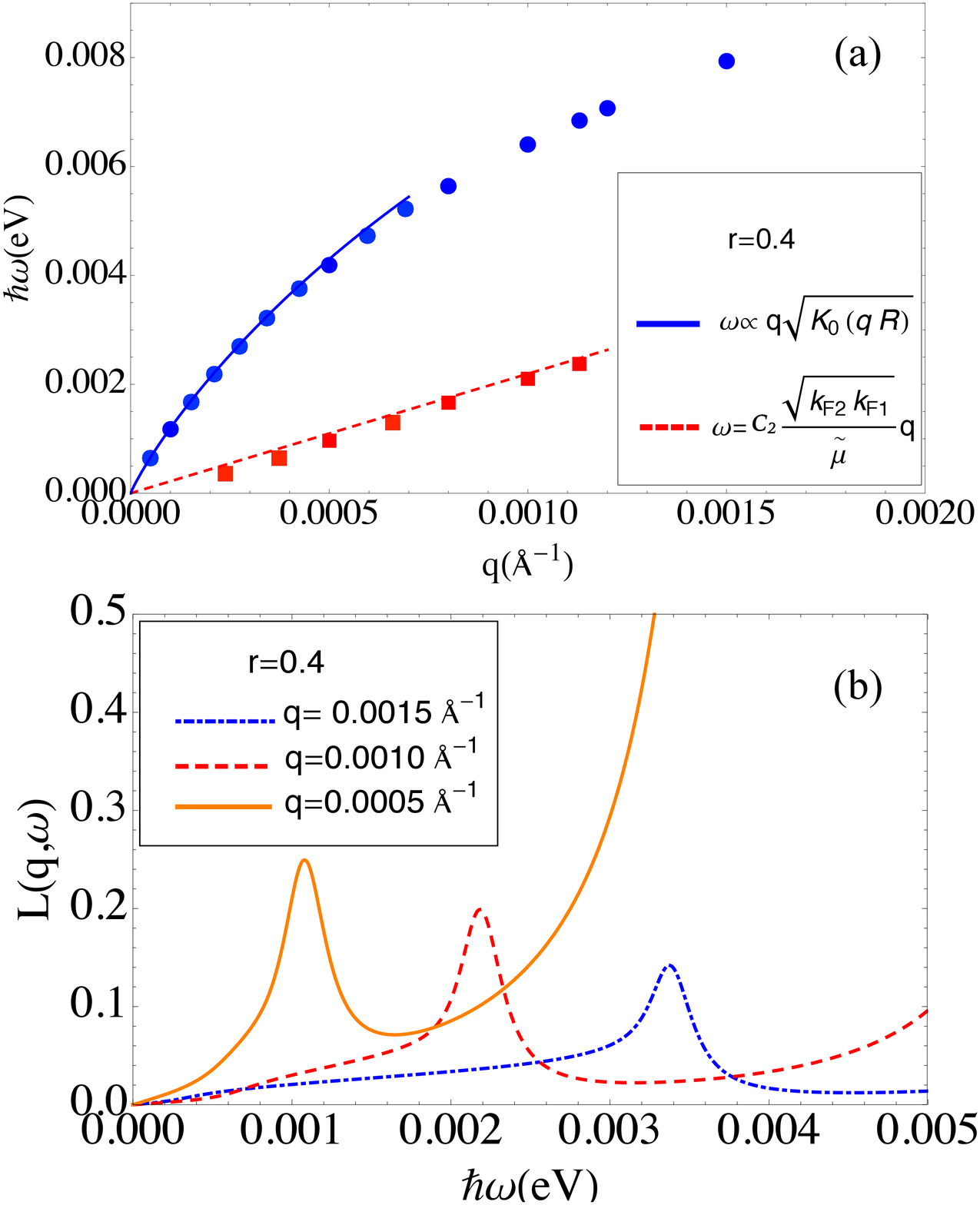}
\caption{ a) Plasmon dispersion (in units of eV) as a function of momentum $q$ (in units of $\AA^{-1}$) for $n=10^{10} \; cm^{-2}$ and $R_0=500 \; \AA$ at $r=0.4$ in the case of two occupied sub-bands. We distinguish a solution for $\omega>C_2q$ (blue circles) and an other with a linear behavior for $\omega<C_2 q$ (red sqares) where $k_{F1}=k_F(m=0)$ and $k_{F2}=k_F(m=-1)$. Continuous lines indicate the analytical solutions. b) The dynamic structure factor as a function of the energy (in units of eV, energy range $\hbar \omega<C_2 q$) for different values of wave-vector $q$ at $r=0.4$ in the case of two occupied sub-bands with  $n=10^{10} \; cm^{-2}$.}
 \label{fig:6}
 \end{figure}

If we apply a magnetic field parallel to the axis of the cylinder, we remove the degeneration of  energy sub-bands. Keeping constant the electron doping to $n=10^{10} \; cm^{-2}$ for $R_0=500 \AA$, inevitably the chemical potential changes crossing more sub-bands. This depends on the ratio  $r$  of the fluxes as seen in equation (\ref{eq:43}). Therefore, if we take, for example, a value of  $r=0.4$, the chemical potential crosses two sub-bands instead of one as before. Even in this case, by exploiting Eq.(\ref{eq:44}), we can get an analytical solution of the plasmon dispersion for large wavelength. In fact, in the case of  two sub-bands, in the limit of $q \rightarrow 0$, the relation in (\ref{eq:44-1}) becomes a bi-quadratic equation because we have a sum on $m=0$ and $m=-1$ in  Eq.(\ref{eq:44}). Hence, we get four solutions, two for positive frequencies and two with  opposite frequencies, as shown  in Figure (\ref{fig:6}a) for a specific case. A solution will be of the same type like in the single sub-band case ($\omega \sim q\sqrt{K_0(qR_0)}$), the other will have a linear behavior in $q$ for values $\hbar \omega<C_2 q$. The analytical solution for the linear dispersion provides $\omega=q C_2 \sqrt{k_{F}(0)\;k_{F}(-1)}/\tilde{\mu}$,  although, as shown in Figure (\ref{fig:6}b), this solution have a spectral weight lower than the other for $\hbar \omega>C_2 q$. In fact, for the doping and parameters considered here, the spectral weight of the plasmon peak with linear dispersion is about ten times smaller than the other solution ($\omega \propto q \sqrt{K_0(q R_0)}$) at higher frequency. For example, if we take a $q=0.5 \; m\AA^{-1}$ the spectral weights of the two resulting plasmon peaks are $8 \; meV$ ($\hbar \omega>C_2 q$) and $0.2 \; meV$ ($\hbar \omega< C_2 q$), respectively.

A similar reasoning could be made for any number of sub-bands in the limit $q \rightarrow 0$. However, the analytical solutions of equation (\ref{eq:41}) would become more and more complicated increasing the number of sub-bands. At fixed radius $R_0$, increasing the number of sub-bands means considering higher densities. Also in this case an excitation for $\hbar \omega>C_2 q$ will be observed in the dynamic structure factor and the remaining will be all concentrated in the area $\hbar \omega<C_2 q$ like in the case of two sub-bands. For example, in a case of higher doping, for example $5 \times10^{10} \; cm^{-2}$ for $R_0=500 \AA$, the chemical potential crosses three sub-bands. We get a plasmon excitation with a behavior of type $\omega_p\propto q\sqrt{K_0(qR_0)}$  and the other two solutions are both linear in $q$ at lower frequencies. More the electronic doping increases, more the spectral weight of excitations present at $\hbar \omega<C_2 q$ decreases. In addition, the peak shape observed in the dynamic structure factor for these excitations is no longer a single peak as we see in Figure (\ref{fig:6}b) in the case of two sub-bands. Indeed, the structure becomes a continuous separated by the higher frequency narrow main peak (present for $\hbar \omega>C_2 q$) whose spectral weight gets enhanced with increasing the electron doping. If the electron doping further increases, the spectral weight of the low frequency continuous becomes negligible in comparison with the main peak.

Up to now, in this section, we have presented all the results for a radius $R_0=500 \; \AA$. An increase of the doping requires a larger number $M$ in equation (\ref{eq:43}). When one increases the electron doping up to  $n=10^{12}\; cm^{-2}$ keeping the radius at $R_0=500 \AA$, we take into account a number of sub-bands equal to $M=20$. 
If one increases the radius, one reduces the gap and the sub-bands become more dense. In the case of a larger radius, the effects of the Berry phase are lost since the gap is practically closed and the sub-bands become a continuous. The two-dimensional limit for the electronic surface states is reached  for larger and larger  radii.  In the case of  $R_0=4000 \AA^{-1}$ and high doping about $10^{12} \; cm^{-2}$, we show in Figure (\ref{fig:app}) of Appendix \ref{App:A} that the plasmon dispersion converges to that of the two-dimensional limit \cite{Guinea}.

\begin{figure}[h!]
\centering
\includegraphics[width=8.5cm]{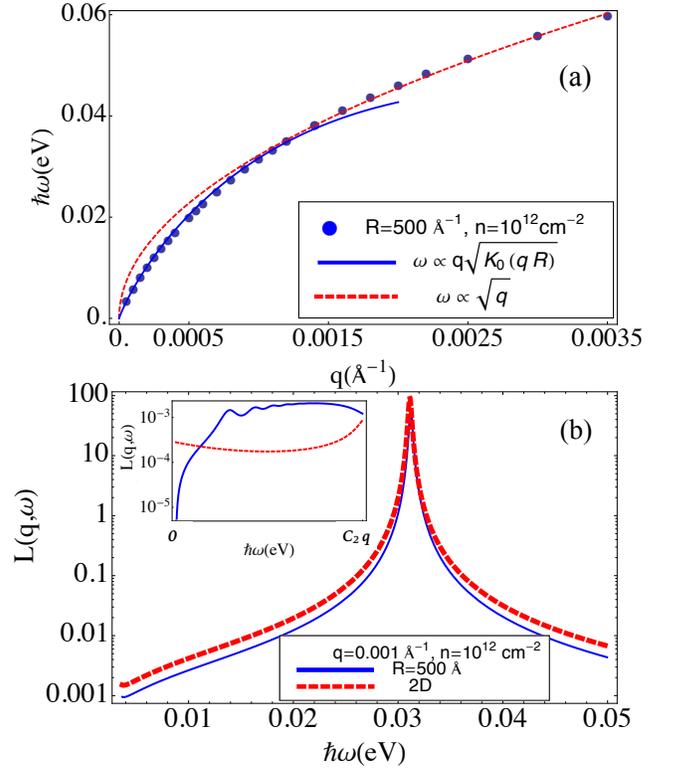}
\caption{ a) Plasmon dispersion (in units of eV, circles) as a function of momentum $q$ (in units of $\AA^{-1}$) for $R_0=500 \AA$ and $n=10^{12} \; cm^{-2}$. For $q<1/R_0$, the dispersion is one-dimensional type: $\hbar \omega = \alpha q\sqrt{K_0(qR_0)}$ (blue solid line) with $\alpha=33 eV \AA$ . For $q>1/R_0$, the dispersion is two-dimensional type: $\hbar \omega= \beta \sqrt{q}$ (red dashed line) with $\beta=1,02 eV \AA^{1/2}$. b) Dynamic structure factor (in logarithmic scale) as a function of the energy (in units of eV) for $q=0.001 \AA^{-1}$, $n=10^{12} \; cm^{-2}$, and $R_0=500 \AA$.  It is compared with the factor of the two-dimensional ($2D$ in figure) infinite surface at the same electron doping and $q$ value. Only in the region $0<\hbar \omega<C_2 q$, for the cylinder case, there is a continuous structure (in the inset), which is not present for the infinite surface.}
 \label{fig:7}
 \end{figure}

On the other hand, for a surface doping  $n=10^{12} \; cm^{-2}$ and at a radius $R_0=500 \AA$, we actually find significant  corrections
 for the plasmon dispersion compared to  the  case of an infinite large cylinder (2D Dirac plasmons). Immediately, we observe a difference in the dispersions for large wavelength. In fact, for $q<1/R_0$, as seen in Figure (\ref{fig:7}a), the dispersion for a radius $R_0=500 \AA$ has one-dimensional features of type $\hbar \omega =\alpha q \sqrt{K_0(q R_0)}$, with $\alpha \sim 33$ $eV \AA$. For small $q$ but such that $q>1/R_0$, we have the two-dimensional behavior: $\hbar \omega = \beta \sqrt{q}$, with $\beta \sim 1.02$ $eV \AA^{1/2}$. A clear dimensional crossover occurs for intra-band plasmons in these range of parameters.
 As discussed in Appendix \ref{App:A}, in the case of a radius $R=4000 \AA$ for the same doping, one gets a two-dimensional dispersion  for all momenta $q$: $\hbar \omega= \beta_2 \sqrt{q}$, with $\beta_2=\sqrt{(C_2 e^2k^{2D}_F)/2}$ and $k^{2D}_F$ two-dimensional Fermi momentum.   Note that, for the chosen parameters,  one gets 
 $\beta_2 \sim 0.98$ $eV \AA^{1/2}$. Therefore, the $\beta$ coefficient is slightly different from what is found for a Dirac 2D plasmon and depending on the radius. 

We notice that the spectral weights for $R_0=500 \AA$ and $R=4000 \AA$ are practically identical as it is possible to observe for a particular $q$ in Figure (\ref{fig:7}b). {The only tiny difference is seen in the region $0<\hbar \omega<C_2 q$. In fact, in the case of 2D Dirac plasmon,  no structure is  observed in that region except for that due to the finite $\eta$ used for the calculation giving a little imaginary part to the spectrum.} For the case of radius $R_0=500 \AA$ instead we see a small structure with a spectral weight that is five orders of magnitude smaller than the weight of the solution for $\hbar \omega>C_2 q$, as shown in the inset of 
Figure (\ref{fig:7}b). Finally, for values of doping larger than $10^{12} \; cm^{-2}$, the plasmon is very stable and does not undergo significant variations even upon the application of magnetic field. Therefore, the Berry phase  does not affect the features of the magneto-plasmon in the limit of a large radius or high doping.

\begin{figure}[h!]
\centering
\includegraphics[width=8.5cm]{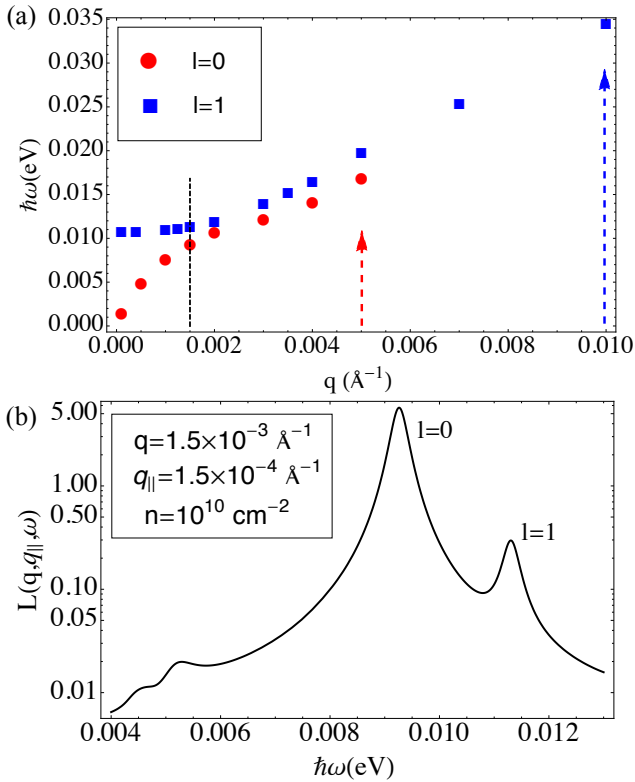}
\caption{ a) Energy (in units of eV) of intra-band plasmons as a function of wave-vector $q$ (in units of $ \AA^{-1}$) at $l=0$ and $l=\pm 1$ for doping $n=10^{10} \, cm^{-2}$ and radius $R_0=500 \, \AA$. The upper cut-off of $q$ is shown for $l=0$ (red dashed arrow) and $l=\pm 1$ (blue dashed arrow).  b) Dynamic factor structure as a function of the energy (in units of eV) at $q=1.5 \times 10^{-3} \, \AA^{-1}$ (black dashed line in (a)) and $q_{||}=1.5 \times 10^{-4} \, \AA^{-1} $. }
 \label{disp_l_intra}
 \end{figure}

Finally, in analogy with the case of zero surface doping, we find the dispersions of intra-band plasmon-like excitations at $l \ne 0$ for finite dopings. In Figure (\ref{disp_l_intra}) we focus reported only on the value $l = 1$ (which is equal to $l=-1$)  for a doping $n=10^{10} \, cm^{-2}$ (single occupied sub-band) and for $R_0=500 \, \AA$. 
Unlike the case of the inter-band plasmon-like excitations, we see immediately from the Figure (\ref{disp_l_intra}a) that the dispersion for $l> 0$ changes a lot in comparison with that for $l=0$. In fact, we find for $l = 1$ (analogously to higher $l$) a plasmon mode whose frequency is higher than that of $l=0$ and constant for small $q$. This plasmon dispersion at $l \ne 0$ is typical of cylindrical metallic wires \cite{sarid}. Even in the case of the intra-band plasmon a cut-off exists for the wave-vector. However, for intra-band plasmons, the plasmon solution is found up to a maximum value of $q$ both for $l=0$ (red dashed arrow in Figure (\ref{disp_l_intra}a)) and for $l=\pm1$ (blue dashed arrow in Figure (\ref{disp_l_intra}a)). 

We can calculate the spectral weights for all the values of $l$. To this aim, we focus on values of $q_{||}$ which are an order of magnitude smaller than $q$. This way, the momentum in nearly parallel to the cylinder axis. Moreover, small values of $q_{||}$ ensures convergence of the series in $l$ by taking a few terms at finite $l$. For a particular value of $q_{||}$, we have analyzed the contributions for $l> 0$ in comparison with that at $l=0$. In general, the term at  $l=0$ is always the most important contribution to dynamic structure factor. In Fig. (\ref{disp_l_intra}b), by using a logarithmic scale, we show the total dynamic structure factor, which is essentially given by the sum of the contribution of $l=0$ and $l=\pm1$. We note immediately that,  for a fixed $q$, the spectral weight associated with $l=0$ is an order of magnitude larger than the weight related to $l=1$. So we confirm that for $q_{||}\ne 0$ but small, the relevant contribution is $l=0$. Another feature that emerges from the Figure (\ref{disp_l_intra}b) is  the presence of a small peak at a frequency lower than the peak at $l=0$. This low frequency small peak corresponds to a  pole that we find from RPA equation always for $l=\pm1$. This solution presents a minimum and maximum $q$ for which the dispersion exists. Of course this peak is about two orders of magnitude smaller than the main peak and tends quickly to reduce. 

\section{Comparison between inter- and intra-band plasmon features}

In this paper, up to now, for intra- and inter-band excitations, we have investigated the dispersion and the damping through the dynamic structure factor. In this section, we clarify the substantial differences that exist between the two modes. By the comparison of the dynamic structure factor  in Figures (\ref{fig:3}a) (inter-band case) and (\ref{fig:7}b) (intra-band case), we notice that the damping of the  inter-band plasmon-like excitation is much larger than that of intra-band modes. In fact, it is possible to observe that the full width at half maximum (FWHM) $\Gamma$ of the dynamic structure factor at a fixed wave-vector $q$ is larger than that of intra- band modes. In order to better analyze the effects of the losses on the mode propagation, we study the behavior of FWHM $\Gamma$ as a function of $q$ for both zero doping and finite doping ($n=10^{12} \, cm^{-2}$) in correspondence with a radius $R_0=500 \, \AA$. 

\begin{figure}[h!]
\centering
\includegraphics[width=8.5cm]{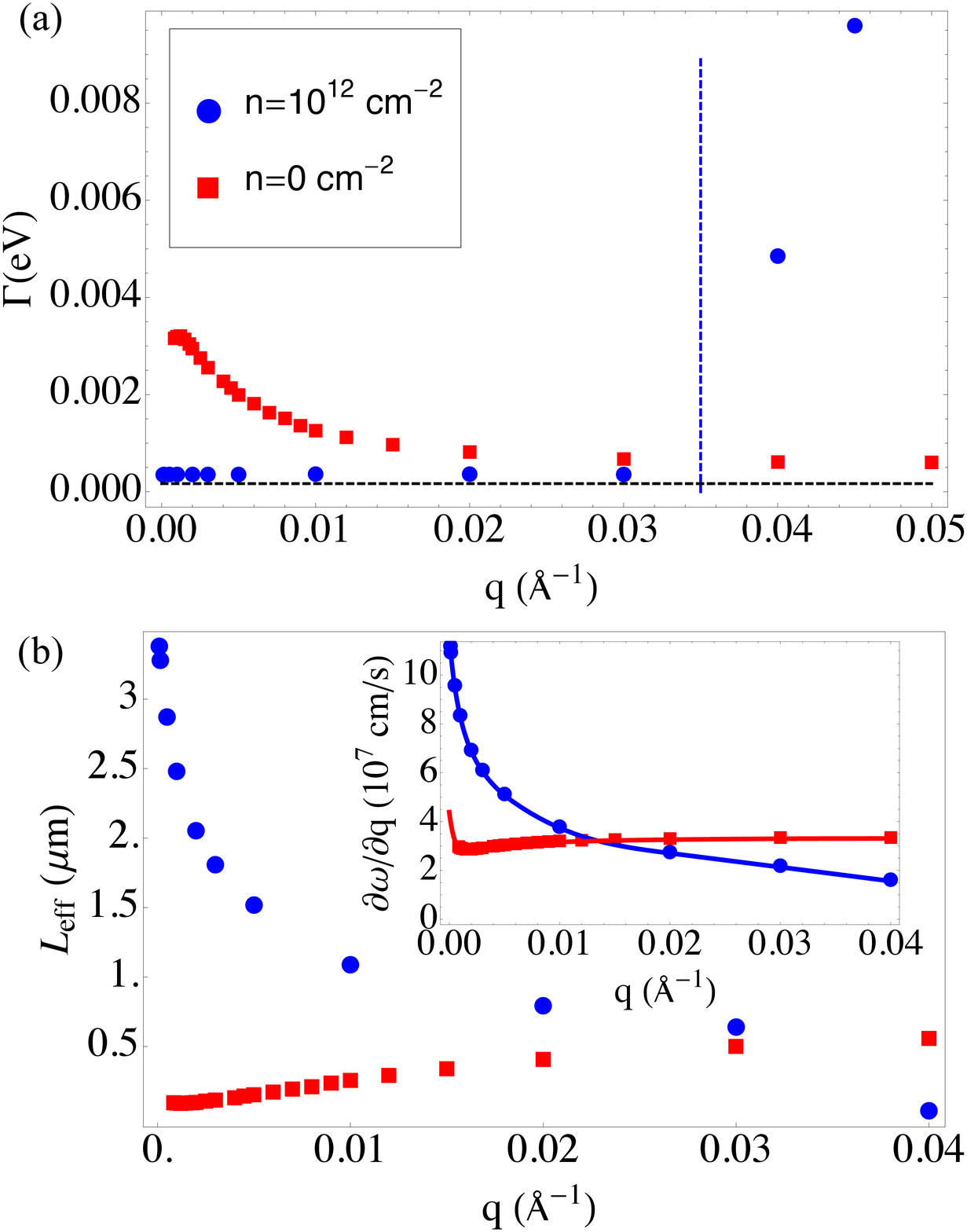}
\caption{ a) Full width at half maximum $\Gamma$ (in units of eV) as a function of $q$ (in units of $\AA^{-1}$) for inter- (zero surface doping, red squares) and intra-band (surface doping $n=10^{12} \, cm^{-2}$, blue circles) plasmons in correspondence with a  radius $R_0=500 \, \AA$. The surface doping $n=10^{12} \, cm^{-2}$ corresponds to a Fermi wave-vector $k_F=0.035 \AA^{-1}$ (dashed blue line). Note as the damping of intra-band plasmon is close to $\eta$ (dashed black line) for $q<k_F$.   b) Propagation distance (in units of $\mu m$) for inter- (zero surface doping, red squares) and intra-band (surface doping $n=10^{12} \, cm^{-2}$, blue circles) plasmons as a function of $q$ (in units of $\AA^{-1}$)  in correspondence with a  radius $R_0=500 \, \AA$. In the inset, the velocity (in units of $10^7$ cm/s) as a function of $q$ (in units of $\AA^{-1}$) for inter- (zero surface doping, red squares) and intra-band (surface doping $n=10^{12} \, cm^{-2}$, blue circles) plasmons.  }
 \label{Gamm_inter_intra}
 \end{figure}

It is possible to observe in Figure (\ref{Gamm_inter_intra}a) that the  FWHM $\Gamma$ associated with the intra-band  plasmon is zero for values of $q<k_F$, where $k_F$ is indicated by a blue dashed line. In fact, the FWHM $\Gamma$ appears to be of the same order of the parameter $\eta$ indicated as a black dashed line in Figure (\ref{Gamm_inter_intra}a) ($\eta$ is equal to $1.5 \times 10^{-4} eV$ in this paper). We have checked that, if we send $\eta$ to zero, then also $\Gamma$ of intra-band plasmons goes to zero  for $q<k_F$. On the other hand, again for intra-band plasmons, we notice a rapid increase of $\Gamma$ for $q>k_F$. This behavior of the intra-band FWHM is very similar to that observed in the case of a two-dimensional system \cite{Guinea}. Therefore,  in the case of infinite radius, we recover not only the plasmon dispersion plasmon but also the behavior of the damping associated with it. 

A different situation occurs for inter-band plasmon-like excitations. As shown in Figure (\ref{Gamm_inter_intra}a), $\Gamma$ for inter-band plasmons is different from zero for small values of the wavevector $q$. Indeed, the value of $\Gamma$ is at least one order of magnitude larger than $\eta$. Considering the peak values of inter-band plasmons shown in  Figures (\ref{fig:3}a), we point out that, for $q=0.001 \, \AA$, $\Gamma$ is about $3 \times 10^{-3} eV$, while the peak energy is around $10^{-2} eV$. Therefore, for  
$q=0.001 \, \AA$,  $\Gamma$ is about one third of the peak energy indicating that the inter-band plasmon is severely damped. We remark that intra-band plasmons have a zero $\Gamma$ for the same values of wave-vectors. Moreover, again in contrast with the intra-band case, $\Gamma$ for inter-band plasmons goes to zero with increasing $q$.  We stress that the decrease of $\Gamma$ with increasing $q$ does not mean that the inter-band plasmon is more defined. In fact, we have checked that the spectral weight related to the peaks of the dynamic structure factor rapidly decreases  with increasing $q$. Actually, for $q=0.05 \, \AA$, the spectral weight can be assumed as  virtually zero.  This behavior is again different when compared to the intra-band case.\\

In the inset of Figure (\ref{Gamm_inter_intra}b), we report the plasmon group  velocity $ \partial \omega/ \partial q$ (in units of $10^7$ $cm/s$) derived from plasmon dispersions. Inter-band and intra-band plasmon modes show different behaviors as a function of $q$. In fact, the intra-band velocity is monotonically decreasing, while the inter-band plasmon is quite constant with increasing $q$. Actually, inter-band velocity slowly increases reaching a plateau, that converges to the asymptotic value equivalent to $C_2/\hbar$. In fact, at zero doping, for enough large $q$, the plasmon dispersion flattens on the value equal to $C_2 q$. We conclude from the features of the velocity that the nature of the inter-band plasmon is profoundly different from the intra-band one.

Finally, it is also possible to calculate the plasmon propagation length $L_{eff}$, which is so defined: $L_{eff}=  \Gamma^{-1} \partial \hbar \omega/ \partial q$. Of course, in addition to the average life time ($\hbar \Gamma^{-1}$), this distance is also related to the group velocity of the plasmon modes. In Fig. (\ref{Gamm_inter_intra}b), we show the propagation length associated with different values of $q$ in the case of inter-band and intra-band plasmons. It is possible to observe how the  intra-band plasmon, in the range where it has a value of $\Gamma$ close to zero, has a length greater than the inter-band one. 
For example, for  $q=0.001 \, \AA$, $L_{eff}$ is of the order of $3 \mu m$ for intra-band plasmon, while it if of the order of $0.2 \mu$m for the inter-band plasmon. Therefore, intra-band plasmons show lengths slightly smaller than those in metals, whose surface plasmons propagate on distances of the  order of ten $\mu$m \cite{sarid}.  On the other hand, inter-band plasmons are characterized by small axial propagation lengths comparable with the radius of the cylinder. The intra-band and inter-band lengths have opposite behaviors as a function of the wave-vector $q$. With increasing $q$, the intra-band length decreases very fast, while the inter-band length slowly increases. In the intra-band case, only for $q>k_F$, the lengths become comparable with inter-band ones.    \\


\section{Conclusions}

In this paper we have studied the charge response of a cylindrical wire of the TI $Bi_2Se_3$. We have provided an analytical solution of the dynamic structure factor within the RPA approximation for the electron-electron Coulomb interaction. We have analyzed how the Berry phase and the interference on it due to an axial magnetic field affect the plasmon dispersion and spectral weights of a TI cylinder. 

One of the main results of the work is that, in the case of zero surface doping ($\mu=0$),  inter-band plasmon-like excitations can form in TI nanowires  unlike in TI slabs. In particular, a strong dependence of this class of excitations on an axial magnetic field has been observed due to the interference effects between the Berry phase and magnetic field. Actually, an axial magnetic field induces a change in dispersion and peak shape of the magneto-plasmon, which becomes more defined with a longer life-time.

Then, we have considered the case of a low surface electron doping of about $10^{10} \; cm^{-2}$ when the chemical potential crosses one sub-band in the absence of magnetic field. We have stressed that the resulting plasmon excitation has both dispersion and spectral weight similar to those of a magneto-plasmon at zero surface doping. In this case, it is has been also possible to calculate the limit for $q \rightarrow 0$ of the polarization function analytically solving the RPA equation and providing the dispersion of intra-band plasmons. The application of a magnetic field removes the degeneration of sub-bands giving rise to two types of excitations: one for $\omega> C_2 q$  with a behavior $\omega \propto q \sqrt{ {\tilde V}_q}$, and another for $\omega< C_2q$ with a linear behavior as a function of $q$. The most important difference between these two kinds of solution is in the spectral weight which is much larger for plasmons at $\omega> C_2q$. 

Finally, we have checked that the limit of infinite two-dimensional surface is reached for a very large radius, for example $R_0=4000 \AA$,  at  a large surface doping of $10^{12} \; cm^{-2}$. We have also analyzed corrective terms to the case of the two-dimensional limit in the plasmon dispersion. To this aim, we have considered the same surface doping of $10^{12} \; cm^{-2}$ but with a smaller radius equal to $R_0=500 \AA$. In any case, for high doping, we have pointed out that the effects induced by an axial magnetic field are no more important. We point out that quite large changes of doping have been already achieved with a gate geometry in TI \cite{Sessi}, in particular in thin films \cite{Ngabonziza}. Therefore, we believe that it is not impossible to tune experimentally the doping in TI nanowires from small to large values.

Focus of this paper has been on the case when, in the dynamic structure factor, the momentum ${\mathbf q_{||}}$  parallel to $x-y$ plane (orthogonal to the axis $z$ of the cylinder) is zero.  We have also studied the effects of ${\mathbf q_{||}} \ne 0$ analyzing charge excitations where the angular degrees of freedom are more involved.
It could be interesting to consider not only the presence of Dirac surface plasmons but also of other plasmon excitations \cite{Talebi}, in particular of surface plasmons \cite{Mingda} related to bulk charge carriers which have been observed very recently in EELS experiments \cite{Kogar,Politano}. Finally, it could be useful to study plasmon features at finite temperatures analyzing the effects of  electron-phonon couplings \cite{Autore1,Cataud}, in particular, in the case of wires freely suspended or grown on a substrate \cite{Nocera}. Work in this direction is in progress.

\section*{Acknowledgements}
C.A.P. acknowledges partial financial support from the Progetto Premiale CNR/INFN EOS Organic Electronics for Innovative
Research Instrumentation. C.A.P. and V.C. acknowledge partial financial support from the regione Campania project L.R.N.5/2007 Role of interfaces in magnetic strongly correlated
oxides: manganite heterostructures.

\begin{appendix}

\section{Calculation and large radius limit of the polarization function and plasmon dispersion}
\label{App:A}

As discussed in section 2, we consider the surface wave functions of a TI cylindrical wire obtained from previous works \cite{Paolino,Imura}:
\begin{equation}
\label{eq:10}
\psi_{k,m}^{s}(\rho,\phi,z) =\sqrt{\frac{1}{4 \pi L}}e^{i k z}R(\rho) e^{i m \phi}\mathbf{u}_{k,m}^{s}(\phi),
\end{equation}
where  $L$ is the length along the $z$ axis, $R(\rho)$ is the radial function, $\mathbf{u}_{k,m}^{s}(\phi)$ is a quadri-spinor depending on the angular variable $\phi$. In Eq.(\ref{eq:10}), the radial function $R(\rho)$  is zero for $\rho \ge R_0$, while, for  $0<\rho<R_0$, it is the following:
\begin{equation}
\label{eq:radial}
R(\rho)=N e^{-\alpha_{-}R_0}\left(e^{(\alpha_{+}+\alpha_{-})\rho} -e^{\alpha_{+}R_0+\alpha_{-}\rho} \right),
\end{equation}
with $N$ normalization constant, 
\[
\alpha_{\pm}=\frac{1\pm\sqrt{1+4a}}{2L_0},
\]
\[
a=-\frac{M_2M_0}{A^2}, \;
L_0=\frac{M_2}{A}.
\]

In Eq.(\ref{eq:10}), for negative eigenvalues $\epsilon^-_{k,m}$, the quadri-spinor is

\begin{equation}
\mathbf{u}^-_{k,m}(\phi)=
\begin{bmatrix} 
\label{eq:11-1}
   \,  \frac{1}{2}+\alpha_{k,m} +i \beta_{k,m} \, \\ 
   \,  -\frac{i}{2}+i(\alpha_{k,m} +i \beta_{k,m} )   \,     \\
   \, -\frac{e^{i \phi }}{2}+e^{i \phi } (\alpha_{k,m} +i \beta_{k,m} ) \,     \\
   \,  \frac{1}{2} i e^{i \phi }+i e^{i \phi } (\alpha_{k,m} +i \beta_{k,m} ) \,
\end{bmatrix},
\end{equation}
while, for positive eigenvalues $\epsilon^+_{k,m}$, it is
\begin{equation}
\label{eq:11-2}
\mathbf{u}^+_{k,m}(\phi)=
\begin{bmatrix} 
   \,  \frac{1}{2}-(\alpha_{k,m} +i \beta_{k,m} ) \, \\ 
   \, -\frac{i}{2}-i(\alpha_{k,m} +i \beta_{k,m} )   \,     \\
   \, -\frac{e^{i \phi }}{2}-e^{i \phi } (\alpha_{k,m} +i \beta_{k,m} )  \,     \\
   \, \frac{1}{2} i e^{i \phi }-i e^{i \phi } (\alpha_{k,m} +i \beta_{k,m} )    \,
\end{bmatrix}.
\end{equation}
with
 \begin{eqnarray}
 \label{eq:11}
 \alpha_{k,m} &=&\frac{\left(\frac{1}{2}+m- r\right) C_1} {2 \sqrt{C_2^2 \text{k}^2+\left(\frac{1}{2}+m- r \right)^2 C_1^2}}, \\
 \beta_{k,m} &=&\frac{C_2 k} {2 \sqrt{C_2^2 \text{k}^2+\left(\frac{1}{2}+m- r \right)^2 C_1^2}}.
 \end{eqnarray} 

We can evaluate the term $\chi_0(z-z^\prime,\phi-\phi^\prime;\omega)$ present in the equation (\ref{eq:4}) for the full  susceptibility:
\begin{equation}
\begin{split}
\label{eq:13}
&\chi_0(z-z^\prime,\phi-\phi^\prime;\omega)= \frac{1}{(2 \pi L)^2}\sum_{k,q,m,l,s,s^\prime} e^{i q (z-z^\prime)}\\
&\times \, e^{i l (\phi-\phi^\prime)} \frac{f(\epsilon^{s}_{k,m} ) - f(\epsilon^{s^\prime}_{k+q,m+l})} {\epsilon^s_{k,m}-\epsilon^{s^\prime}_{k+q,m+l}+\hbar(\omega+i0^+)} F^{s,s^\prime}(k,q,m,l),  
\end{split}
\end{equation}
wheree  $q$ is the momentum along $z$ axis, $l$ is the angular number, and $F^{s,s^\prime}(k,q,m,m+l)$ is the scalar product between the eigenvectors in (\ref{eq:11-1}) and (\ref{eq:11-2}):
\begin{eqnarray}
\label{eq:15}
&&F^{s,s^\prime}(k,q,m,m+l)=\frac{1}{2}\biggl[1+ ss^\prime \times \\ \nonumber
&&  \frac{\tilde{\Delta}^2(R_0) \left(1+2m+2l\right)\left(1+2m\right)+k(k+q)}{\sqrt{\tilde{\Delta}^2(R_0) (1+2m)+k^2}\sqrt{\tilde{\Delta}^2(R_0) (1+2m+2l)+(k+q)^2}} \biggr],
\end{eqnarray}
with
\[
\tilde{\Delta}(R_0)=\frac{C_1}{2 C_2 R_0}.
\]
Finally, transforming in the equation (\ref{eq:13}) the discrete sum over $k$ in a integral, we get the polarization function $\chi_0(q,l;\omega)$ used in the main text:
\begin{equation}
\begin{split}
\label{eq:14}
\chi_0(q,l;\omega)=\frac{1}{(2\pi)^2}\int dk & \sum_{m,s,s^\prime} \frac{f(\epsilon^{s}_{k,m} ) - f(\epsilon^{s^\prime}_{k+q,m+l})} {\epsilon^s_{k,m}-\epsilon^{s^\prime}_{k+q,m+l}+\hbar(\omega+i0^+)} \\
&\times F^{s,s^\prime}(k,q,m,l).
\end{split}
\end{equation}
The polarization function in equation (\ref{eq:14}) is expressed in terms of a sum on $m$. We will show that, in the limit of infinite radius, the double sum in $k$ and $m$ will transform in a bi-dimensional integral over ${\bf k}$. Therefore, equation (\ref{eq:14}) becomes the susceptibility of a 2D problem in the large radius limit. To this aim,  for simplicity, we take only the contributions for $l=0$. The energy for our problem is defined as $\epsilon_{k,m}=\pm C_2 \sqrt{k^2+ (m+1/2)^2/R_0^2}$ having considered $C_1=C_2$. In the 2D case we have $\epsilon_{k}=\pm \hbar v_F \sqrt{k^2+ k_{||}^2}$, where $C_2=\hbar v_F$,  $k$ is the same associated to $z$ direction and ${\bf k}_{||}$ is the other momentum in the plane. If we send $R_0 \rightarrow \infty$ also the parameter $m \rightarrow \infty$, and, if $k_{||}=\frac{2 \pi}{L}m$, where $L=2 \pi R_0$, then in this limit $m/R_0 \rightarrow k_{||}$. The factor $1/2$ due to Berry phase, in the limit of large radius, is no longer relevant. Automatically, even the pre-factors $F^{s,s^\prime}(k,q,m,m+l)$ becomes the two-dimensional ones:
 \begin{equation}
\label{eq:16}
F^{s,s^\prime}(k,|\bold{k}_{||}|,q) =\frac{1}{2}\left[1+ss^\prime \frac{ k (k+q)+ k_{||}^2}{\sqrt{k^2+k_{||}^2} \sqrt{(k+q)^2+k_{||}^2}}\right].
 \end{equation}  

Finally, the discrete sum on $m$ becomes a continuum such that $\sum_{m} \rightarrow R_0 \int dk_{||}$. Naturally, we note that between $\chi_0(q;\omega)$ and 
$\chi_{2D}(q;\omega)$ there is a length factor of difference, therefore, we conclude that
\begin{equation}
\label{eq:17}
\frac{1}{R_0}\chi_0(q;\omega) \xrightarrow{R_0 \rightarrow \infty} \chi_{2D}(q;\omega)
\end{equation}
where $\chi_{2D}(q;\omega)$ is
\begin{equation}
\begin{split}
\label{eq:18}
\chi_{2D}(q,\omega)=\frac{1}{(2\pi)^2}\int d{\bf k} & \sum_{s,s^\prime} \frac{f(\epsilon^{s}_{k} ) - f(\epsilon^{s^\prime}_{k+q})} {\epsilon^s_{k}-\epsilon^{s^\prime}_{k+q}+\hbar(\omega+i0^+)} \\\times &F^{s,s^\prime}(k,k_{||},q).
\end{split}
\end{equation}

If we also considered contributions for $l \ne 0$, in addition to the momentum $q$ another transferred momentum $q_{||}$ of transversal type to the axis of the cylinder would appear. 

We can find through the electron density function, the equation of the RPA theory given by $1- {\tilde V}_{q,l}\chi_0(q,l;\omega)=0$, which provides the condition for plasmon formation in the system. For this reason we need to expand the Coulomb potential in the cylindrical coordinates  as follows:
\begin{equation}
\begin{split}
V(\vec{r}-\vec{r^\prime})=\frac{2}{\pi}\sum_{m=-\infty}^\infty\int_0^\infty dk&
I_{m}(k r_<)K_{m}(k r_>)\\
&\times e^{i m(\phi-\phi^\prime)}\cos[k (z-z^\prime)]
\label{eq:19}
\end{split}
\end{equation}
which, by using translational invariance in the $z$ direction, becomes 
\begin{equation}
V(q;{\bf r},{\bf r^\prime})=2\sum_{m=-\infty}^\infty
I_{m}(|q|r_<)K_{m }(|q| r_>)e^{i m (\phi-\phi^\prime)}
\label{eq:20}
\end{equation}
where we used
\begin{equation}
\begin{split}
\int dz e^{-iq(z-z^\prime)}&\cos[k (z-z^{\prime})]=\\
&=2\pi\frac{1}{2}
\left[\delta(q-k)+\delta(q+k) \right].
\label{eq:21}
\end{split}
\end{equation}

From the definition, the induced density can be written as $d(\vec{ r})=\int d\vec{ r^\prime} d\vec{r^{\prime\prime}} \chi(\vec{ r},\vec{r^\prime}) V(\vec{r^\prime}-\vec{r^{\prime\prime}}) d(\vec{r^{\prime\prime}})$ where $\chi(\vec{r},\vec{r^\prime};\omega) =R(\rho)^2R(\rho^\prime)^2  \chi_0(z-z^\prime,\phi-\phi^\prime;\omega)$ is the susceptibility. Passing in Fourier transform in $q$ and in $l$, and  replacing the equation (\ref{eq:13}) and (\ref{eq:20}) in $d(\vec{ r})$,  we have that the density is given by:
\begin{eqnarray}
d_{q,l}(r)&=& 4 \pi e^2 \chi_0(q,l)\int d \rho^\prime \rho^\prime R(\rho)^2R(\rho^\prime)^2\nonumber\\
&\times&\int d\rho^{\prime\prime}\rho^{\prime\prime} 
I_n(|q|\rho_<)K_n(|q|\rho_>)\cdot d_{q,l}(\rho^{\prime\prime}).
\label{eq:22}
 \end{eqnarray}

If we define:
  \begin{eqnarray}
  \label{eq:23}
  S_{q,l}(\rho^{\prime\prime})&=&\int d\rho^\prime \rho^\prime R(\rho^\prime)^2I_l(|q|\rho_<)K_l(|q|\rho_>)\\
  &=&K_l(|q| \rho^{\prime\prime})\int_0^{\rho^{\prime\prime}} d\rho^\prime \rho^\prime R(\rho^\prime)^2I_l(|q|\rho^\prime)\nonumber \\
  &+&I_l(|q| \rho^{\prime\prime})\int_{\rho^{\prime\prime}}^{R_0} d\rho^\prime \rho^\prime R(\rho^\prime)^2K_l(|q|\rho^\prime) , \nonumber
\end{eqnarray}
and multiplying by $S_{q,l}(\rho)$ to the left and right member and integrating over $\rho$, the equation in (\ref{eq:22}) becomes:
 \begin{equation}
 \label{eq:24}
 1=   \chi_0(q,l) {\tilde V}_{q,l}.
  \end{equation}
where we called $ {\tilde V}_{q,l}=4 \pi e^2 \int d\rho \rho R(\rho)^2S_{q,l}(\rho)$.

\begin{figure}[h!]
\centering
\includegraphics[width=8.5cm]{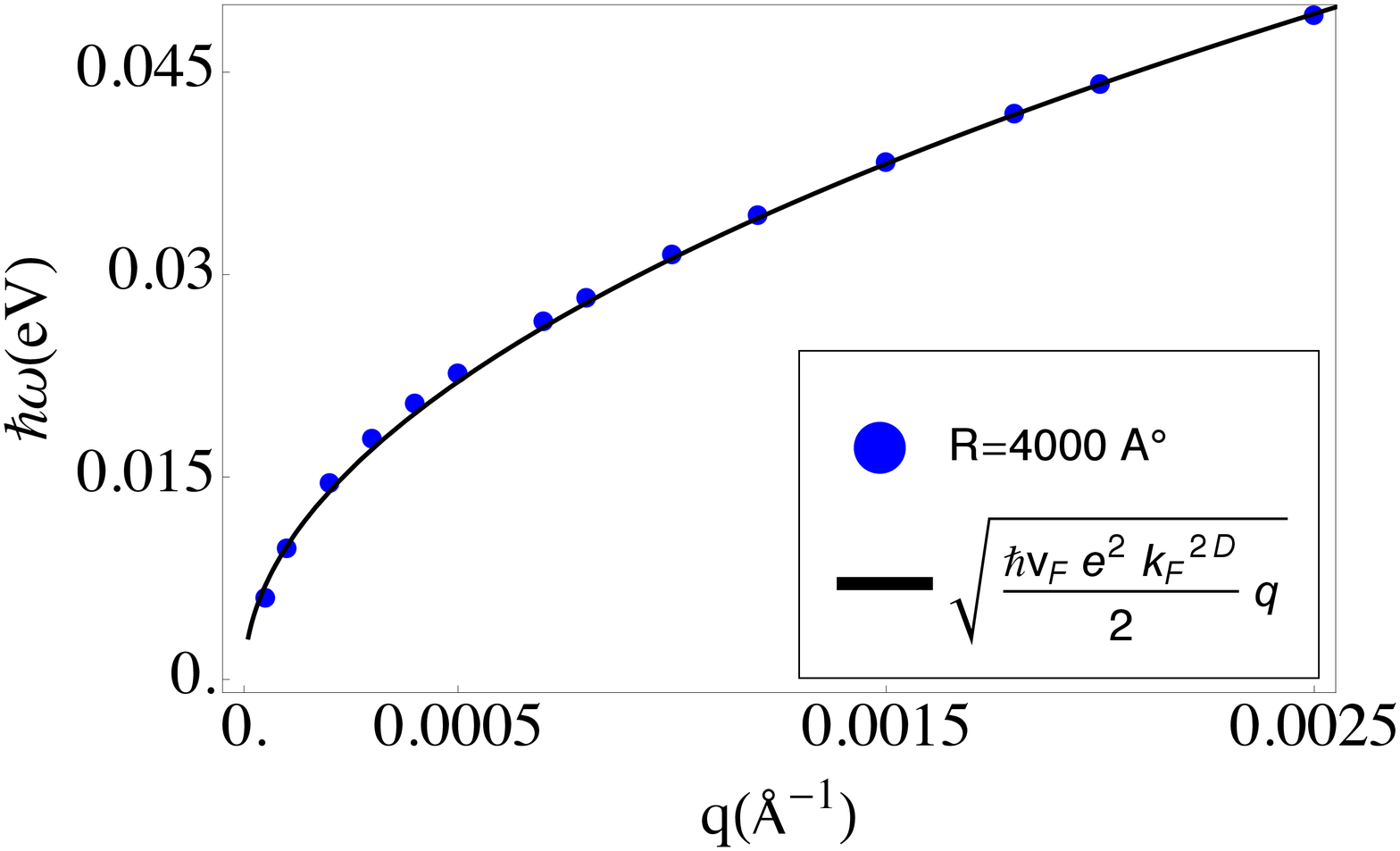}
\caption{\label{fig:app} The plasmon dispersion as a function of the momentum $q$ along $z$ for a cylindrical wire of radius $R=4000 \AA$ . The result is compared with that of an infinite plane (translational invariance along all two directions \cite{Guinea}).  In this limiting case, there is an analytical form for the plasmon dispersion which is reported in the figure.}
 \end{figure}

Recalling that $l=0$, we note that, if $R_0$ is very large, for long wavelength $q R_0 \ll 1$, thus we can write
\begin{eqnarray}
\label{eq:25}
R_0  {\tilde V}_{q,l=0} =4 \pi e^2R_0&&\int d\rho \rho R(\rho)^2S_{q,0}(\rho)  \\\nonumber
 && \mapsto 4\pi e^2 R_0\; I_0(q R_0)K_0(q R_0) \mapsto \frac{2\pi e^2}{ q},
\end{eqnarray}  
which is the transformed of the two-dimensional Coulomb potential. Therefore the equation in (\ref{eq:24}) for large $R_0$ becomes 
\[
1= \frac{1}{R_0} \chi_0(q,0) \frac{2 \pi e^2}{q}
\]
where, in this limit, from equation (\ref{eq:17}), $\chi_0(q,0)/R_0$ converges towards the two-dimensional susceptibility. As shown in the Figure (\ref{fig:app}), the plasmon dispersion of the infinite radius nanowire converges to that of a 2D problem ($\omega \propto \sqrt{q}$).

As discussed in section III, we can derive the induced density and potential in correspondence with the plasmon frequencies. Upon substituting Eq. (\ref{eq:24}) in Eq. (\ref{eq:22}), one immediately finds that, in our approach, the induced density corresponding to a mode with given $q$ and $l$ is the following:
\begin{equation}
 d(\rho,z,\phi)=B_{q,l}(R_0) R^2(\rho) e^{i q z}e^{i l \phi} ,
 \label{rhotot}
\end{equation}
where $B_{q,l}(R_0)$ is an arbitrary constant dependent not only on $q$ and $l$, but also on the cylinder radius $R_0$. As expected, the density depends on the radial function $R^2(\rho)$, which, for a sufficiently large radius, is peaked practically only on the lateral surface of the cylinder.  

From the density, one can determine the electrical potential $\Phi$ corresponding to a plasmon with given $q$ and $l$:
\begin{eqnarray}
\label{eq:poten1}
\Phi(\rho,z,\phi)&= & e^{i q z}e^{i l \phi} B_{q,l}(R_0) \times  \nonumber \\ 
&& \int_{0}^{R_0} d\rho^{\prime}\rho^{\prime} R^2(\rho^{\prime}) K_l(q \rho_>) I_l(q \rho_<).
\end{eqnarray}

Eq. (\ref{eq:poten1}) can be specialized to the case $\rho<R_0$, yielding $\Phi^{int}$ (potential internal to the cylinder) 
\begin{eqnarray}
\label{eq:poten2}
\Phi^{int}(\rho,z,\phi)&= & e^{i q z}e^{i l \phi} B_{q,l}(R_0) \times \\  \nonumber
&& \bigg[ K_l(q \rho) \int_{0}^{\rho} d\rho^{\prime}\rho^{\prime} R^2(\rho^{\prime})  I_l(q \rho^\prime) +  \\ \nonumber 
&& I_l(q \rho) \int_{\rho}^{R_0} d\rho^{\prime}\rho^{\prime} R^2(\rho^{\prime})  K_l(q \rho^\prime) \bigg],
\end{eqnarray}
and to case $\rho>R_0$, providing $\Phi^{ext}$ (potential external to the cylinder) 
\begin{eqnarray}
\label{eq:poten3}
\Phi^{ext}(\rho,z,\phi)&= & e^{i q z}e^{i l \phi} B_{q,l}(R_0) \times \\  \nonumber
&&  K_l(q \rho) \int_{0}^{R_0} d\rho^{\prime}\rho^{\prime} R^2(\rho^{\prime})  I_l(q \rho^\prime).
\end{eqnarray}
Recalling that the associated electric field is given by $\vec{E}(\vec{r})=-\nabla \Phi(\vec{r})$, we can derive the cylindrical components $\left(E_\rho, E_z, E_{\phi}\right)$ of the field in correspondence with a plasmon mode. We stress that for $l=0$ the contribution $E_{\phi}=0$ in all the space, so that, for this mode, only the longitudinal and radial components of the field survive. Therefore, the l=0 mode bears strong similarity with the TM mode found  in correspondence with surface plasmons of metallic cylinders only for $l=0$  \cite{sarid}. In our case, we have analyzed the electrostatic limit, therefore the axial component of the magnetic field is clearly zero.

We observe that, for the radii analyzed in this paper ($R_0 \ge 500 \AA$), the radial wavefunction present in Eqs. (\ref{eq:poten2}) and (\ref{eq:poten3}) can be approximated as a delta function: $R^2(\rho) \simeq  \delta(\rho-R_0)/R_0$. Clearly, this approximation becomes better and better with increasing the radius of the cylinder. 
Therefore, the induced density $\rho$ becomes different from zero only on the lateral surface whose charge density is proportional to $1/R_0$. 
We can provide analytical expressions for the electric field focusing in particular on the modes with $l=0$: $\left(E_\rho,E_z \right)= B_{q,l=0}(R_0) \times$
\begin{equation}
\label{fielddd}
 \left\{
        \begin{array}{ll}
          \left(q I_1(q \rho) K_0(q R_0) e^{i q z}, \, i q  I_0(q \rho) K_0(q R_0) e^{i q z} \right) , \ \  \rho \le R_0, \\
           \left( -q K_1(q \rho) I_0(q R_0) e^{i q z}, i q K_0(q \rho) I_0(q R_0) e^{i q z} \right) , \ \  \rho >R_0.
        \end{array}
    \right.
\end{equation}
We point out that the longitudinal component $E_z$ is continuous at $\rho=R_0$, while the radial component $E_\rho$ is discontinuous. Actually, using the Wronskian of the modified Bessel functions of zero order \cite{sarid}, one can prove that the difference $E_{\rho}(\rho \rightarrow R_0^+)- E_{\rho}(\rho \rightarrow R_0^-)$ is proportional to $1/R_0$, therefore to the charge surface density relative the lateral surface of the cylinder. Inside the cylinder ($\rho<R_0$),  the  radial component $E_\rho$ goes as $ I_1(q \rho)$, therefore, at finite $q$, 
$E_\rho$ is proportional to $q \rho$ for very small values of $\rho$.  On the other hand, outside the cylinder ($\rho>R_0$), $E_\rho$ goes as $K_1(q \rho)$, therefore, at finite $q$, 
$E_\rho$ is proportional to $ e^{-q \rho}/ \sqrt{q \rho} $ for very large values of $\rho$.   

It is interesting to determine the behavior of the electric field in different limits.  For large wavelengths (small $q$, with $q \ll 1/R_0$), the field in Eq. (\ref{fielddd}) becomes
\begin{equation}
\label{eqq}
 \left\{
        \begin{array}{llll}
           E_{\rho} \propto \frac{q^2 \rho}{2} K_0(q R_0) , \ \ & \quad \rho \le R_0, \\ 
           E_{\rho} \propto -\frac{1}{\rho}  I_0(q R_0) , \ \ & \quad \rho>R_0, \\ 
           E_z  \propto i q \left(1+\frac{q^2 \rho^2}{4}\right)  K_0(q R_0), \ \ &   \quad \rho \le R_0, \\ 
           E_z  \propto -i q \log{(q \rho)}  I_0(q R_0), \ \ & \quad R_0<\rho<1/q,
        \end{array}
    \right.
\end{equation}
recalling that the modified Bessel functions can be written as $I_0(q \rho) \sim (1+q^2 \rho^2/4)$ and $K_0(q \rho) \sim -\log(q \rho)$ for $q \rho \ll 1$. 
In particular, from Eq. (\ref{eqq}) emerges that, for $q \rightarrow 0$,  as expected, the electric field inside the cylinder ($\rho<R_0$) is zero. In the same limit $q \rightarrow 0$, outside the cylinder ($\rho>R_0$),  the longitudinal component $E_z$ vanishes, while the radial component $E_\rho$ goes as $1/\rho$. We can also determine the electric field in the limit of small wavelengths (large $q$, with $q>1/R_0$):
\begin{equation}
\label{e1qq}
 \left\{
        \begin{array}{llll}
           E_\rho \propto \frac{1}{\sqrt{2 \pi q}} e^{q \rho}  \frac{(2 q \rho-1)}{2 \rho^{3/2}} K_0(q R_0) , \ \ & \quad \rho \le R_0, \\ 
           E_\rho \propto -\sqrt{\frac{\pi}{2 q}} e^{-q \rho}  \frac{(2 q \rho+1)}{2 \rho^{3/2}} I_0(q R_0) , \ \ & \quad \rho>R_0, \\ 
           E_z \propto i q \frac{1}{\sqrt{2 \pi q \rho}}e^{q \rho}  K_0(q R_0), \ \ &   \quad \rho \le R_0, \\ 
           E_z \propto i q \frac{\sqrt{\pi}}{\sqrt{2  q \rho}} e^{-q \rho} I_0(q R_0), \ \ & \quad \rho>R_0,
        \end{array}
    \right.
\end{equation}
recalling that the Bessel functions can be written as $I_0(q \rho) \sim e^{q \rho}/\sqrt{2 \pi q \rho}$ and $K_0(q \rho) \sim e^{-q \rho}\sqrt{\pi/2 q \rho} $ for $q \rho > 1$.

\section{Calculation of the inverse dielectric function and dynamic structure factor}
\label{App:B}
In this appendix we calculate the inverse of the dielectric constant and, then, the dynamic structure factor. This calculation could be extended to any type of geometry, although here for simplicity it is specialized for a cylindrical case. Exploiting the symmetries of the cylinder and using equation (\ref{eq:19}), the dielectric constant given in equation (\ref{eq:6}) can be written as:
 \begin{equation}
 \begin{split}
 \label{eq:30}
 &\epsilon(\vec{r},\vec{r^\prime})= \frac{\delta(\rho-\rho^\prime)}{\rho} \delta(\phi-\phi^\prime) \delta(z-z^\prime)  \\ 
 & -  \frac{2}{ L}   R(\rho^\prime)^2 \sum_{q_z} e^{i q_z (z-z^\prime)} \sum_l e^{i l (\phi-\phi^\prime)}  S_{q_z,l}(\rho) \chi_0(q_z,l) .
 \end{split}
  \end{equation}

 The inverse dielectric function is so defined:
 \begin{equation}
 \label{eq:31}
 \int d\vec{r_1} \epsilon^{-1}(\vec{r},\vec{r_1}) \epsilon(\vec{r_1},\vec{r^\prime})=\delta(\vec{r}-\vec{r^\prime}).
 \end{equation}
Using $q$ and $l$ and replacing the equation (\ref{eq:30}) in (\ref{eq:31}), we get :
  \begin{eqnarray}
  \label{eq:32}
 &\int d\rho_1 \rho_1&\epsilon^{-1}_{q,l}(\rho,\rho_1) \biggl( \frac{\delta(\rho_1-\rho^\prime)}{\rho_1} + \\ \nonumber
&& -4 \pi  R(\rho^\prime)^2  S_{q,l}(\rho_1) \chi_0(q,l) \biggr)= \frac{\delta(\rho-\rho^\prime)}{\rho} .
\end{eqnarray}

We need to solve an integral equation with a separable variable kernel.  If we define  $A_{q_,l}(\rho)=\int d\rho_1 \rho_1\epsilon^{-1}_{q,l}(\rho,\rho_1) S_{q,l}(\rho_1) $ and we multiply for $S_{q,l}(\rho^\prime)$, integrating both members over $\rho^\prime$, the integral equation becomes: 
 \begin{equation}
\label{eq:33}
 A_{q,l}(\rho) \left[1- 4 \pi  \chi_0(q,l) \int d\rho^\prime \rho^\prime  R(\rho^\prime)^2  S_{q,l}(\rho^\prime) \right]= S_{q,l}(\rho).
\end{equation}
We note that the equation in square parenthesis is equal to equation in (\ref{eq:24}) since $ {\tilde V}_{q,l}=4 \pi \int d\rho^\prime \rho^\prime  R(\rho^\prime)^2  S_{q,l}(\rho^\prime)$. This way, it is possible find the term $A_{q,l}$ from the equation (\ref{eq:33}) that replaced in (\ref{eq:32}) returns
\begin{eqnarray}
\label{eq:34}
\epsilon^{-1}_{q,l}(\rho,\rho^\prime)= \frac{\delta(\rho-\rho^\prime)}{\rho} + \frac{S_{q,l}(\rho)\chi_0(q,l) R(\rho^\prime)^2}{1-\chi_0(q,l) _{q,l}} .
 \end{eqnarray}
 
Finally, we can write a dynamic structure factor in this form:
\begin{equation}
\label{eq:35}
L(\omega,q,q_{||})=-Im\left[\frac{1}{V}\int \int d\vec{r} d\vec{r^\prime} \epsilon^{-1}(\vec{r},\vec{r^\prime}) e^{-i \vec{q} \cdot \vec{r}} e^{i \vec{q} \cdot \vec{r^\prime}} \right],
\end{equation}
where $V$ is the volume enclosed by the cylinder. Exploiting the cylindrical symmetries and passing in Fourier transform with $ \epsilon^{-1}(\vec{r},\vec{r^\prime})= \frac{1}{L} \sum_{q} e^{i q (z-z^\prime)} \frac{1}{2 \pi} \sum_{l} e^{i l (\phi-\phi^\prime)} \epsilon^{-1}_{q,l}(\rho,\rho^\prime) $, we get for the only integral part over $z$ and $z^\prime$ in (\ref{eq:35}):
\begin{eqnarray}
\label{eq:36}
\frac{1}{L^2}\frac{L}{2 \pi}\int dq \int \int dz dz^\prime e^{i(q-q)z} e^{i(q-q)z^\prime}= 1.
\end{eqnarray}

Finally, we have:
\begin{eqnarray}
\label{eq:37}
L(\omega,q,q_{||}) &=& -\frac{1}{2 \pi^2 R_0^2}Im \biggl[ \sum_l \frac{\chi_0(q_z,l;\omega)}{1-\chi_0(q,l;\omega)  {\tilde V}_{q,l}} \\ \nonumber
&& \times \int d\rho^\prime \rho^\prime d\phi^\prime e^{i |{q_{||}}|  \rho^\prime \cos{\phi^\prime}} e^{-i l \phi^\prime} R(\rho^\prime)^2 \\ \nonumber
&&\times \int d\rho \rho d\phi e^{i |{q_{||}}|  \rho \cos{\phi}} e^{i l \phi} S_{q,l}(\rho) \biggr] 
\end{eqnarray}
where both the integrals in $\phi$ and $\phi^\prime$ return the Bessel functions $4 \pi^2 {\rm J}_l(|q_{||}|\rho){\rm J}_l(|q_{||}|\rho^\prime)$. The integral in (\ref{eq:37}) is thus obtained:
\begin{eqnarray}
\label{eq:38}
&& L(q,q_{||}; \omega)= -\frac{2}{R_0^2} Im \biggl [ \sum_l  \frac{\chi_0(q,l;\omega)}{1-\chi_0(q,l;\omega)  {\tilde V}_{q,l}} \\
&&\times  \int_{0}^{R_0} d\rho^\prime \rho^\prime \,  {\rm J}_l(|q_{||}|\rho^\prime) R(\rho^\prime)^2 \int_0^{\infty} d\rho \rho\, {\rm J}_l(|q_{||}|\rho) S_{q,l}(\rho)\biggr]. \nonumber
\end{eqnarray}
 
We notice that the second integral in equation (\ref{eq:38}) depends on $S_{q,l}$ which is linked to the Coulomb potential. Since the Coulomb interaction between electrons is effective in the whole space, inside and outside the material, the integration limits of the radial variables are extended from zero to infinity.

\end{appendix}

\end{document}